\documentclass[prb,twocolumn,floatfix,showpacs,superscriptaddress]{revtex4-1}

\usepackage{epsfig}
\usepackage{amsmath}
\usepackage{amssymb}

\newcommand{\defgradtensor}{\mathbf{F}}

\newcommand{\Ctensor}{\mathbf{C}}
\newcommand{\ctensor}{\mathbf{c}}
\newcommand{\Xvector}{\mathbf{X}}
\newcommand{\xvector}{\mathbf{x}}
\newcommand{\uvector}{\mathbf{u}}
\newcommand{\pfcepsilon}{\varepsilon}

\begin{document}

\title{Non-linear elastic effects in phase field crystal and amplitude equations: Comparison to ab initio simulations of bcc metals and graphene}

\date{\today}

\author{C. H\"uter}
\affiliation{Institute for Energy and Climate Research, Forschungszentrum J\"ulich GmbH, D-52425 J\"ulich, Germany}
\affiliation{Max-Planck-Institut f\"ur Eisenforschung GmbH, D-40237 D\"usseldorf, Germany}
\author{M. Fri\'ak}
\affiliation{Institute of Physics of Materials, Academy of Sciences of the Czech Republic, v.v.i., \\ \v{Z}i\v{z}kova 22, CZ-616 62 Brno, Czech Republic}
\affiliation{Central European Institute of Technology, CEITEC MU, Masaryk University, Kamenice 5, CZ-625 00 Brno, Czech Republic}
\author{M. Weikamp}
\affiliation{Max-Planck-Institut f\"ur Eisenforschung GmbH, D-40237 D\"usseldorf, Germany}
\author{J. Neugebauer}
\affiliation{Max-Planck-Institut f\"ur Eisenforschung GmbH, D-40237 D\"usseldorf, Germany}
\author{N. Goldenfeld}
\affiliation{Department of Physics, Loomis Laboratory of Physics, University of Illinois at Urbana-Champaign, 1110 West Green Street, Urbana, Illinois 61801-3080, USA}
\author{B. Svendsen}
\affiliation{Chair of Material Mechanics, RWTH Aachen University, 52062 Aachen, Germany}
\author{R. Spatschek}
\affiliation{Institute for Energy and Climate Research, Forschungszentrum J\"ulich GmbH, D-52425 J\"ulich, Germany}

\begin{abstract}
We investigate non-linear elastic deformations in the phase field crystal model and derived amplitude equations formulations.
Two sources of non-linearity are found, one of them based on geometric non-linearity expressed through a finite strain tensor.
It reflects the Eulerian structure of the continuum models and correctly describes the strain dependence of the stiffness.
In general, the relevant strain tensor is related to the left Cauchy-Green deformation tensor. 
In isotropic one- and two-dimensional situations the elastic energy can be expressed equivalently through the right deformation tensor.
The predicted isotropic low temperature non-linear elastic effects are directly related to the Birch-Murnaghan equation of state with bulk modulus derivative $K'=4$ for bcc. 
A two-dimensional generalization suggests $K'_{2D}=5$.
These predictions are in agreement with ab initio results for large strain bulk deformations of various bcc elements and graphene.
Physical non-linearity arises if the strain dependence of the density wave amplitudes is taken into account and leads to elastic weakening.
For anisotropic deformations the magnitudes of the amplitudes depend on their relative orientation to the applied strain.

\end{abstract}


\maketitle

\section{Introduction}



For the understanding and development of new materials with specific mechanical properties, a good knowledge of the elastic response is mandatory.
A complete parametrisation of elastic properties either experimentally or via {\em ab initio} techniques is however challenging, especially if information beyond the linear elastic regime is required, which is important for high-strength materials.
Whereas in the linear elastic regime the number of elastic constants is limited, it is obvious that a complete characterization of the mechanical response in the non-linear regime increases the number of required parameters tremendously.
A reduction of this parameter set, together with an increased understanding for the non-linear behaviour, would therefore be beneficial.
Therefore, the present paper aims at a reduction of this complexity by exploiting the intrinsic description of non-linear elasticity provided by the phase field crystal (PFC) model, in combination with {\em ab initio} calculations as well as analytical energy-volume relations.   

The PFC method\cite{Elder:2004ys, PhysRevLett.88.245701} has become a  popular method for simulating microstructure evolution on diffusive timescales and with atomic resolution.
In contrast to conventional phase field models, this approach allows to describe e.g.~the detailed structure of grain boundaries, as the atomic density distribution is maintained.
The PFC community has extended the scope of the model tremendously over the years, and we just mention few of the recent remarkable developments here. Hydrodynamics have been included \cite{HeinonenV:2016prl}, as well as growth from vapor phases \cite{KocherG:2015prl}, dislocation dynamics \cite{TarpJM:2014prl}, and glass formation \cite{BerryJ:2011prl}, and recently also polycrystalline 2D materials, in particular graphene \cite{MSeymour:2016aa}. 
General structural transformations became accessible by constructing free energy functionals from generic two-particle correlation functions \cite{GreenwoodM:2010prl}. 

One of the advantages of the PFC model is that it automatically contains elasticity, as a deformation of the lattice, expressed through a change of the lattice constant, raises the energy.
For small deformations this energy change is quadratic, hence linear elasticity is captured, and for larger deformations non-linear effects appear\cite{Chan:2009aa, Huter:2015aa}.
Whereas the original PFC model is fully phenomenological, later extensions have shown that it can be linked to the classical density functional theory of freezing \cite{Singh1991351, ISI:A1987K583500046, Spatschek:2010fk}, which allows to determine the model parameters from fundamental physical quantities, which can for example be determined from molecular dynamics simulations\cite{Wu:2006uq,Wu:2007kx,Adland:2013ys,kar13}.
The obtained elastic constants can then be obtained from properties of the liquid structure factor and give reasonable estimates of the high temperature values near the melting point.
Conceptually, this means that the theory is applied in the high temperature regime, formally at the coexistence between solid and melt phases.
Here, in contrast, we aim at an understanding of the ability of the model to capture elastic effects also in the low temperature regime.
The predictions will be compared to {\em ab initio} results using electronic structure density functional theory (DFT). We note that for the elastic constants and non-linearity the reliability of DFT calculations is excellent, as it is supported by experimental benchmarks\cite{LejaeghereK:2013crit}. 

The article is organized as follows:
In Section \ref{ModelingApproachSection} we revisit the ingredients for the work in the present article.
It starts with general concepts concerning the description of non-linear elasticity and then discusses them in the context of the PFC model.
The section is concluded with details on the {\em ab initio} simulations, which are used to benchmark the continuum descriptions.
Section \ref{1DPFCsection} analyses the non-linear elastic response of the PFC model in one dimension, where the analysis is particularly simple, emphasizing the Eulerian character of the elastic description and the different roles of geometric and physical non-linearity.
Section \ref{2DPFCsection} continues with the two dimensional situation of crystals with triangular symmetry.
Section \ref{sectionBCC} investigates the same behavior for the three-dimensional case of bcc crystal structures.
The results are compared to classical descriptions of non-linear elasticity in Section \ref{BirchMurnaghanSection} and also to {\em ab initio} simulations of bcc materials and graphene.
The article concludes with a summary and discussion in Section \ref{SummaryConclusion}.

\section{Modeling approach}
\label{ModelingApproachSection}

It is one of the primary goals of the present article to link expressions for the elastic response under large deformations using modeling approaches on different scales.
For small deformations, the energy increases quadratically with the strain, as known from linear theory of elasticity.
For larger deformations, deviations appear, which require a careful distinction between the undeformed reference and the present state of the material.
We first investigate these non-linear effects from a mechanical perspective, which we then apply to the phase field crystal model.
Here, a primary goal is to see which of the different non-linear strain tensors is most suitable to describe the elastic response in these models.
The results are compared to {\em ab initio} simulations of large bulk deformations for various bcc metals and graphene.

\subsection{Finite strain description}

We start the investigations with a brief reminder and definition of the different finite strain tensors and quantities relevant for mechanical applications and modelling as described e.g.~in Refs.~\onlinecite{LandauLifshitz:7,NematNasserHori:1,DimitrienkoY:2011b}. 

We begin with the deformation gradient tensor $\defgradtensor$ which describes how a medium at a reference point $\Xvector$ is deformed when it changes to a new coodinate $\Xvector \mapsto \xvector$. 
The local geometry is represented by a segment $d\Xvector$ in the reference  system which is mapped to $d\xvector$ in the deformed system.
We define $dx_j = F_{jk} dX_k$ in coordinate representation, using the Einstein convention for summation over repeated indices. 
We note that intuitively the deformation gradient tensor can be constructed as product of a rotation tensor and and a stretch tensor, or, more generally a non-rotational tensor. 
For the description in terms of finite displacements, we introduce the relative displacement vector $d\uvector$ which describes the difference between the two segments $d\Xvector,d\xvector$ as $d\uvector = d\xvector-d\Xvector$ in addition to the displacement of the reference point from $\Xvector$ as $\uvector = \xvector-\Xvector$. 
Then the right Cauchy-Green deformation tensor $\Ctensor= \defgradtensor^\dagger \defgradtensor$ (with $\dagger$ for the transposition) describes the square of local change of distances by deformation as $d\xvector^2 = d\Xvector \Ctensor  d\Xvector$. In coordinate representation, it is connected to the Green-Lagrange finite strain tensor $\epsilon_{ij}$ components as 
\begin{equation} \label{LagrangianStrain}
\epsilon_{ij} =\frac{1}{2} \left( C_{ij} - I_{ij} \right) = \frac{1}{2} \left( \frac{\partial u_i}{\partial X_j} + \frac{\partial u_j}{\partial X_i} + \frac{\partial u_m}{\partial X_i}\frac{\partial u_m}{\partial X_j}  \right).
\end{equation}
The corresponding tensor in the reference frame of the deformed medium is the Piola tensor $\ctensor = \defgradtensor^{-1\dagger}\defgradtensor^{-1}$, which is related to the Euler-Almansi finite strain tensor as
\begin{equation}\label{eulerStrain}
e_{ij} =\frac{1}{2} \left( I_{ij} - c_{ij} \right) = \frac{1}{2} \left( \frac{\partial u_i}{\partial x_j} + \frac{\partial u_j}{\partial x_i} - \frac{\partial u_m}{\partial x_i}\frac{\partial u_m}{\partial x_j}  \right).
\end{equation} 
Finally, we introduce the tensor 
\begin{equation} \label{strangeEulerStrain}
\bar{e}_{ij} = \frac{1}{2}\left( I_{ij} - C^{-1}_{ij}\right) =\frac{1}{2} \left( \frac{\partial u_i}{\partial x_j} +  \frac{\partial u_j}{\partial x_i}  - \frac{\partial u_i}{\partial x_k}\frac{\partial u_j}{\partial x_k} \right),
\end{equation}
corresponding to the $D$ strain of Clayton \cite{Clayton:2014aa}.
It involves the inverse of the right Cauchy-Green deformation tensor $C^{-1}_{ij}=(\delta_{ik}-\partial u_i/\partial x_k)(\delta_{jk}-\partial u_j/\partial x_k)$. 
Here, the differences between the tensors (\ref{eulerStrain}) and (\ref{strangeEulerStrain}) appear as contraction either in the numerator or denominator of the non-linear part.
The above finite strain tensors all agree up to the level of leading terms (as used in linear elasticity), and deviations due to the geometric effects show up at the quadratic level.
For a recent discussion of the three different strain tensors from a continuum mechanics perspective of large deformations we refer to Ref.~\onlinecite{Clayton:2014aa}.

In the following, we use the term {\em geometric non-linearity} to express the fact the elastic energy depends on finite deformation measures (e.g., stretch or strain).
These will be identified below.
Due to the non-linear terms in the strain tensors the energy is therefore not a quadratic function in terms of displacement gradients.
Besides geometric non-linearity,
also {\em physical non-linearity} contributes to deviations from linear elasticity.
Physical non-linearity pertains when the terms in the elastic energy of cubic or higher-order in the (geometrically linear or non-linear) strain become non-negligible.  
A classic example of this is anharmonic elastic behavior. 
It is obvious that such effects should show up at sufficiently large strains.
Under tension, complete dissociation of the material leads to independent atoms or molecules with vanishing interaction and stress.
Under strong compression, the Pauli repulsion leads to stress increases due to ``hard core contributions'', and these effects are not captured by geometric non-linearity alone.

\subsection{Phase field crystal modeling}
\label{pfcintrosec}

The phase field crystal model uses an order parameter $\psi$ to describe a material state.
In contrast to conventional phase field models this order parameter is not spatially constant, but exhibits periodic modulations in a crystalline phase. 

For simplicity, we use here only the original and very basic phase field crystal model, which is described by the energy functional
\begin{equation} \label{pfcfunctional}
F = \int_V d\mathbf{r} \left\{ \psi \left[(q_0^2+\nabla^2)^2 -\pfcepsilon\right]\frac{\psi}{2} +\frac{\psi^4}{4} \right\}. 
\end{equation}
The atom density is denoted by $\psi$, which is periodic in a crystalline state, and $V$ is the system volume.
All quantities are assumed to be dimensionless, and $\pfcepsilon$ is a control parameter, which corresponds to a dimensionless temperature.
In the following we set $q_0=1$. 
We focus here on crystalline phases and ignore the melt phase, in agreement with the concept of a low temperature limit.
The average density $\bar{\psi}$ is a second control parameter.
It is defined as
\begin{equation}
\bar{\psi}=\frac{1}{V} \int_V \psi(\mathbf{r}) d\mathbf{r}.
\end{equation}
The free energy density $f$ in the expression (\ref{pfcfunctional}), i.e.~the expression in curly brackets $\{\cdots\}$, averaged over a unit cell, will later allow the comparison to {\em ab initio} calculated energies.


Equilibrium is obtained via the evolution equation for a conserved order parameter
\begin{equation} \label{PFCevolution}
\frac{\partial\psi}{\partial t} = \nabla^2\left( \frac{\delta F}{\delta\psi}  \right).
\end{equation}
Here we focus on equilibrium elastic properties only, therefore the precise (conserved) dynamics is not important.

\begin{figure}
\begin{center}
\includegraphics[trim=1cm 13.5cm 1.5cm 1cm, clip=true, width=8.5cm]{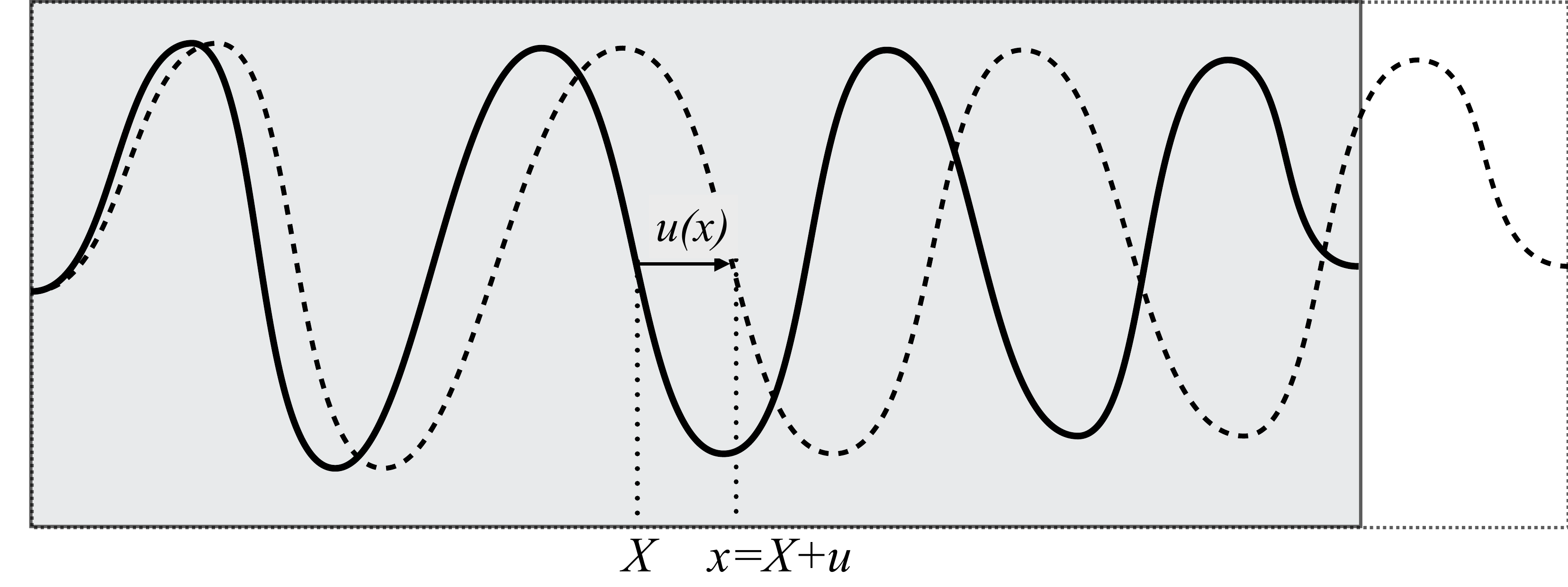}
\caption{One-dimensional sketch of the density profile in the PFC model.
The solid curve is the relaxed reference density, the dashed curve describes a deformed state.
In the PFC model the displacement field $u$ is defined in Eulerian coordinates, hence defined at the deformed coordinate $x=X+u$.}
\label{figsketch}
\end{center}
\end{figure}

A sketch of a density profile in one dimension is shown in Fig.~\ref{figsketch}.
With the basic energy functional given by Eq.~(\ref{pfcfunctional}) different crystal structures like smectic or triangular (in 2D) or bcc (in 3D) are found.
Therefore, the interpretation of the order parameter of an atom density is suggested.
Before starting with the actual analysis it is worthwhile to discuss the interpretation of elasticity within the phase field crystal model, and to become aware of limitations.
Typically, we perform simulations in a fixed volume, as indicated by the grey box in Fig.~\ref{figsketch}.
When the material is deformed, this would physically lead to a change of the system size, which we do not consider in the simulations.
Instead, the domain, in which the equations are solved, is still the grey shaded area.
This already hints at the understanding of elasticity in the PFC model in a Eulerian spirit.
As a result of this fixed system size the number of ``atoms'' -- the peaks in the density profiles -- is not conserved (but see also the discussion about physical atomic density in PFC and vacancies in Ref.~\onlinecite{ChanPY:2009pre}). 
Additionally, the conservation of the particle number can be violated by the creation or annihilation of atoms under large strain (Eckhaus instability\cite{EckhausW:1965b}), and will not be further considered here.

Physically, one would expect from such an interpretation that the density is related to the volume change during deformation.
With the original system size of the undeformed reference state being $V_0$ and the deformed system having volume $V$, one may suggest the relation $\bar{\psi}_0 V_0 = \bar{\psi} V$ as conservation of the average density, which would then change from $\bar{\psi}_0$ to $\bar{\psi}$.
However, such an interpretation is misleading, as instead $\bar{\psi}$ is considered as constant control parameter in the simulations using a fixed volume, and there is no direct connection between the average density $\bar{\psi}$ and the number of atoms.
This is most striking in the one-mode expansion, see Eqs.~(\ref{1Donemode}) and (\ref{onemode}) below, where average density and atom spacing can be varied independently, noting that this approximation gives an excellent description of the true density $\psi$ in particular in the regime of small values of $\pfcepsilon$.
Most interpretations of elasticity in the PFC model use the picture of following the atoms' positions during the deformation, which allows to define the elastic response. 
An exception is the analysis in Ref.~\onlinecite{JaatinenA:2009pre}, which defines the bulk modulus via the density dependence of the free energy.
According to the above discussion it is not surprising that this interpretation leads to different bulk moduli in comparison to the first approach.

Many of the conceptual questions related to the interpretation of elasticity in the PFC model become prominent only when non-linear effects are considered.
In the framework of (geometrically) linear elasticity the difference between Eulerian and Lagrangian strains vanishes, as obvious from the expressions (\ref{LagrangianStrain})-(\ref{strangeEulerStrain}) above, and also the distinction between undeformed and deformed configurations is ignored.

In an atomistic description one considers the energy per unit cell, and by the inspection of this {\em integrated energy} as a function of the strain one can determine the linear and non-linear elastic behavior.
As we will see in the following, one arrives at a physically useful interpretation of elasticity in the PFC model, if one considers the {\em energy density} in a Eulerian sense as measure.
This has implicitly been used in many investigations in the literature \cite{Elder:2004ys, PhysRevLett.88.245701, Spatschek:2010fk} for small deformations, where is it appropriate.
However, a thorough investigation in the non-linear regime is still missing, apart from investigations in Ref.~\onlinecite{Chan:2009aa, Huter:2015aa}.
In particular we find that this interpretation leads to a description analogous to the Birch-Murnaghan equation of state\cite{MurnaghanFD:1944pnas, BirchF:1947pr}, which is frequently used in {\em ab initio} simulations to fit the elastic energy, and also sheds light on the strength of non-linearity for bcc elements and graphene.
This will become more transparent in Section \ref{BirchMurnaghanSection}.

On the practical level we use an analytical description which is based on a one-mode expansion of the density field, as used also to derive amplitude equations descriptions.
This means that we write the density as a superposition of plane waves.
As mentioned before, for small values of $\bar{\psi}$ and $\pfcepsilon$ such a sine wave approximation is very good and allows to treat the problem of non-linear elastic deformation analytically.
This will be shown explicitly in one, two and three dimensions in the following sections, taking care of the important role of geometric non-linearity.
The analysis builds up on the work by Chan and Goldenfeld \cite{Chan:2009aa}, rectifying an improper interpretation of the non-linear strain tensor.

\subsection{Ab initio modeling}

The quantum mechanical calculations within the framework
of density functional theory~\cite{Hohenberg1964, Kohn1965}
are performed using the Vienna Ab Initio Simulation Package
(VASP)~\cite{Kresse1993,Kresse1996}. 
The exchange and correlation
energy is treated in the generalized gradient approximation as
parametrized by Perdew, Burke, and Ernzerhof~\cite{Perdew1996}
and implemented in projector augmented wave
pseudopotentials~\cite{Bloechl1994}. 
We use a plane-wave cutoff
of 450 eV with a 18 $\times$ 18 $\times$ 18 Monkhorst-Pack k-point
mesh for the 2-atom elementary body-centered cubic (bcc) supercells,
yielding total-energy accuracy better than 1 meV per atom.
All calculations are performed at $T=0$.
The corresponding VASP calculations for graphene are performed similarly, 
with a plane-wave cutoff of 350 eV  and a 48 $\times$ 48 $\times$ 3 Monkhorst-Pack k-point mesh for 
the 2-atom hexagonal  cells. 
The computational cells are designed 
to be highly anisotropic in their shapes so as to separate individual 
graphene sheets from their periodic images by 32 \AA{}ngstrom of vacuum.

\section{The one-dimensional PFC model}
\label{1DPFCsection}

We use a one-dimensional situation first to illustrate the conceptual approach.
It briefly summarizes results from the literature\cite{Chan:2009aa} and extends them, elucidating the role of non-linear elasticity. 

\subsection{Geometric non-linearity}

For an analysis of the elastic energy we use a one-mode approximation of the order parameter.
\begin{equation} \label{1Donemode}
\psi(x) = A\sin(q x) +\bar{\psi}
\end{equation}
With this one gets the averaged free energy density\cite{Elder:2004ys} 
\begin{eqnarray} \label{1DPFCelasticEnergy}
f &=& \frac{\bar{\psi}^2}{2} \left[ -\pfcepsilon +1 +\frac{3A^2}{2} + \frac{\bar{\psi}^2}{2} \right] \nonumber \\
&& + \frac{A^2}{4} \left[ -\pfcepsilon + (1-q^2)^2 + \frac{3A^2}{8} \right]. \label{eq3}
\end{eqnarray}
Averaging is done over multiples of the ``unit cell'', i.e.~periods with ``lattice unit'' $a=2\pi/q$.
Obviously, $q=1$ minimises the energy density for fixed amplitude $A$ and $\bar{\psi}$ (for $q_0=1$).
For the moment, we keep the amplitude constant.
Then a variation of $q$ leads to an elastic energy change proportional to $(1-q^2)^2$ for small deviations from the ground state.
A value $q\neq 1$ expresses a homogeneous strain in the system, and therefore the displacement field has the form
\begin{equation} \label{eq2}
u(x) = (1-q) x,
\end{equation}
which turns out to be defined in a Eulerian frame, which will become more obvious below in Section \ref{1DEulerlagrange}.
The displacement gradient is $\partial u/\partial x = 1-q$.
In our present one-dimensional setup we get from the strain definitions (\ref{eulerStrain}) and (\ref{strangeEulerStrain})
\begin{equation} \label{nonumbersofar}
e_{xx} = \bar{e}_{xx} = \frac{1}{2} (1-q^2),
\end{equation}
noting that for a one-dimensional situation the tensors (\ref{eulerStrain}) and (\ref{strangeEulerStrain}) coincide.
Only later, in three dimensional situations we will see that in fact $\bar{e}_{ij}$ is the most suitable tensor in the context of PFC modeling.
Here we see that the elastic energy density can be written in terms of $\bar{e}_{ij}$, as it is proportional to $e_{xx}^2=\bar{e}_{xx}^2$, i.e.~$f_{el}=[f(e_{xx})-f(0)]= A^2\bar{e}_{xx}^2$.
An important result is that the non-linear elasticity on this level can be completely attributed to the {\em geometric} non-linearity.
The constitutive law, which connects stress and strain, is still purely linear, since the elastic energy is quadratic in $\bar{e}_{ij}$.

\begin{figure}
\begin{center}
\includegraphics[width=8.5cm]{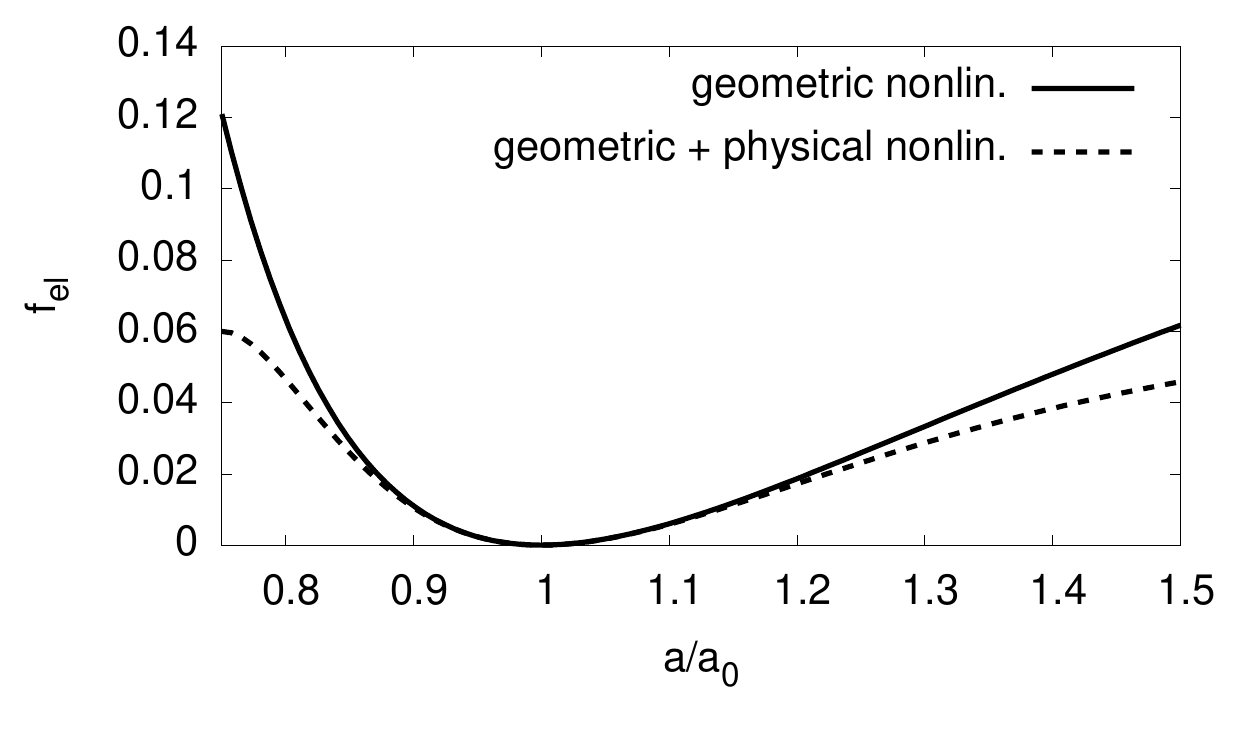}
\caption{
Elastic energy per unit cell as a function of the lattice constant $a$ in the one-dimensional phase field crystal model for $\bar{\psi}=0$ and $\pfcepsilon=0.6$.
The solid curve uses a constant amplitude $A=A(q_0)$, see Eq.~(\ref{eq4}), whereas the dashed curve is based on a strain dependent amplitude $A=A(q)$ according to the same equation.
}
\label{fig2}
\end{center}
\end{figure}

We can plot the elastic energy density (for fixed amplitude) as a function of the lattice constant $a=2\pi/q$, as shown as solid curve in Fig.~\ref{fig2}.
One has to keep in mind that in this representation the energy changes asymmetrically around $a_0=2\pi/q_0$, in contrast to the dependence as a function of the strain $\bar{e}_{xx}$.
It is important to mention that in the non-linear elastic regime the stiffness is higher under compression ($a<a_0$) then under tension, as one would expect physically.

\subsection{Physical non-linearity}

So far we have assumed that the amplitude $A$ is constant and does not depend on the strain, which leads to geometric non-linearity only.
We follow here the analysis by Chan and Goldenfeld \cite{Chan:2009aa} to account for physical non-linearity.

In equilibrium, the value of $A$ is optimised via the condition
\begin{equation}
\frac{\partial f(A, q, \pfcepsilon, \bar{\psi})}{\partial A} = 0.
\end{equation}
From Eq.~(\ref{eq3}) we get
\begin{equation} \label{eq4}
A = \pm 2\left(\frac{\pfcepsilon}{3} - \bar{\psi}^2 - \frac{1}{3} (1-q^2)^2 \right)^{1/2}.
\end{equation}
Close to $q=1$, i.e.~in the linear elastic regime, the amplitude is unaffected by the strain.
For larger deformations, the amplitudes are reduced as a precursor of a strain induced melting process.
Inserting this amplitude back into the energy expression leads to the dashed curve in Fig.~\ref{fig2}. 
For larger strains, the physically non-linear effects overcompensate the geometric non-linearity, as discussed above.
This is more pronounced for smaller values of $\pfcepsilon$, and then we get agreement of the solid and dashed curve essentially only in the linear elastic regime.
This is in line with the interpretation of $\pfcepsilon$ as an undercooling with respect to the solid-liquid coexistence, and therefore for lower values of $\pfcepsilon$ a strain induced melting is more favorable.

We mention that for higher dimensional situations the strain dependence of the amplitudes will be orientation dependent.
Hence, different amplitudes will then depend differently on an anisotropic strain.
This effect has not been considered in Ref.~\onlinecite{Chan:2009aa}.

\subsection{Eulerian vs.~Lagrangian description}
\label{1DEulerlagrange}

Here we demonstrate that a precise distinction of the strain tensors and the reference states is essential for a correct description.
As mentioned before beyond linear elasticity a careful use of deformed and reference configurations is mandatory, and this will be investigated here.

To illustrate this one could naively use the definition (\ref{LagrangianStrain}) to calculate the Lagrangian strain from the displacement (\ref{eq2}), identifying $x$ as the reference coordinates $X$.
This would lead to 
\begin{equation}\label{inconvinientStrain}
\epsilon_{xx}=(1-q)(3-q)/2,
\end{equation}
and the elastic energy contained in (\ref{eq3}) cannot easily be represented through this strain expression.
However, this expression for the strain tensor would be based on an erroneous mixing of reference frames.

In the Lagrangian perspective, one consistently has to work in the undeformed reference state. 
%
For the non-deformed case, we have there a density profile $\psi_0(x) \sim \cos(q_0 x)$ (using $q_0=1$) and for the deformed one $\psi(x) \sim \cos(q x)$, where $x$ is a Eulerian coordinate.
The ``atom'' which is originally located at the (Lagrangian) density peak position $X=2\pi/q_0$ is displaced to $X+u_L(X)=2\pi/q$.
Hence the displacement at the position $X$ is given by
\begin{equation} \label{exampledispl}
u_L(2\pi) = 2\pi/q - 2\pi. 
\end{equation}
With such a homogeneously strained solid the (Lagrangian) deformation gradient is
\begin{equation}
\frac{\partial u_L(X)}{\partial X} = \frac{u_L(2\pi)}{2\pi} = \frac{1}{q}-1,
\end{equation}
since the reference length in the undeformed crystal is $2\pi/q_0$ (notice that the displacement at $x=0$ is zero).
With this, the Lagrangian strain (\ref{LagrangianStrain}) becomes 
\begin{equation}
\epsilon_{xx} = \frac{1}{2} \left( \frac{1}{q^2} -1 \right),
\end{equation}
which clearly differs from the (incorrect) expression in Eq.~(\ref{inconvinientStrain}).
Furthermore, the strain is the relative length change of a material, as expressed through $dx^2=dX^2 + 2 \epsilon_{ij}dX_i dX_j$ in Lagrangian formulation.
Here this leads consistently to $dx^2=dX^2/q^2$, in agreement with the wavelength change.

Let us contrast this to the Eulerian description.
Here the displacement is the same as above in Eq.~(\ref{exampledispl}), but read as a function of the deformed coordinate $x$,
\begin{equation}
u_E(2\pi/q) = 2\pi/q - 2\pi.
\end{equation}
For the inverse deformation gradient the reference is now the deformed system, hence
\begin{equation}
\frac{\partial u_E(x)}{\partial x} = \frac{u_E(2\pi/q)}{2\pi/q} = 1-q.
\end{equation}
Consequently, the Eulerian strain reads according to Eqs.~(\ref{eulerStrain}) and (\ref{strangeEulerStrain}) 
\begin{equation}
e_{xx}=\bar{e}_{xx} = \frac{1}{2}(1-q^2),
\end{equation}
which coincides with Eq.~(\ref{nonumbersofar}) and shows  
that we are indeed operating in a Eulerian description  in the PFC model.
The length change is expressed through the relation $dx^2=dX^2 + 2 e_{ij}dx_idx_j$, which reads here again consistently $dX^2=q^2 dx^2$.


\section{The 2D triangular model}
\label{2DPFCsection}

As in the one-dimensional case we use the amplitude equation formulation to extract the non-linear elastic response of a two-dimensional stable or metastable triangular phase.
The density field is expressed as
\begin{equation} \label{onemode}
\psi = \bar{\psi} + \sum_{j=1}^N \left[ A_j\exp(i\mathbf{k}^{(j)}\cdot\mathbf{r}) + A_j^*\exp(-i\mathbf{k}^{(j)}\cdot\mathbf{r}) \right]
\end{equation}
with $N=3$ here.
The normalised reciprocal lattice vectors (RLVs) are
\begin{equation} \label{2DRLVs}
\mathbf{k}^{(1)} = \left(
\begin{array}{c} 
0\\
1
\end{array}
\right),
\quad 
\mathbf{k}^{(2)} = \left(
\begin{array}{c} 
\sqrt{3}/2\\
-1/2
\end{array}
\right),
\quad
\mathbf{k}^{(3)} = \left(
\begin{array}{c} 
-\sqrt{3}/2\\
-1/2
\end{array}
\right).
\end{equation}
In the following we work in the parameter regime $\bar{\psi}>0$.
With the above definition of the RLVs for an undeformed state the amplitudes are equal in magnitude, but not in sign,
\begin{equation} \label{2Damplitudesigns}
-A_1 = A_2 = A_3= A.
\end{equation}
Chan and Goldenfeld \cite{Chan:2009aa} derived the free energy functional,  which follows from insertion of the amplitude expansion into the free energy functional (\ref{pfcfunctional}) and assuming that the amplitudes vary on a scale which is large in comparison to the atomic spacing.
Then, only terms which correspond to closed polygons of reciprocal lattice vectors contribute, and one arrives at the functional
\begin{eqnarray}
F &=& \int d\mathbf{r} \Bigg\{ -\sum_{j=1}^3 A_j^* (\Gamma-L_j^2)A_j + 3 \sum_{j,\ell=1}^3 |A_j|^2 |A_\ell|^2 \nonumber \\
&& - \frac{3}{2} \sum_{j=1}^3 |A_j|^4 + 6 \bar{\psi} (A_1A_2A_3 +A_1^*A_2^*A_3^*)  \Bigg\} \nonumber \\
&=& \int d\mathbf{r} (f_{local} + f_{nonlocal}), \label{eq22}
\end{eqnarray}
where an offset, which is independent of the amplitudes, is skipped, see Appendix \ref{appendix::aederivation} for details.
This free energy functional contains the operator
\begin{equation}
L_j = \nabla^2 + 2 i \mathbf{k}^{(j)}\cdot\nabla 
\end{equation}
and $\Gamma = \pfcepsilon - 3\bar{\psi}^2$.
The nonlocal term is 
\begin{equation}
f_{nonlocal} = A_j^* L_j^2 A_j. 
\end{equation}
After an integration by part we can represent it more conveniently  as 
\begin{equation}
f_{nonlocal} = |L_j A_j|^2, 
\end{equation}
where we have skipped boundary terms.


For a deformed state the amplitudes are
\begin{equation}
A_j = A_{j, 0} \exp[-i\mathbf{k}^{(j)}\cdot \mathbf{u}(\mathbf{r})],
\end{equation}
which uses the proper sign in the exponential compared to Ref.~\onlinecite{Chan:2009aa} and \onlinecite{Huter:2015aa}.
This is in line with the above discussion in Section \ref{1DEulerlagrange} and Ref.~\onlinecite{Spatschek:2010fk}.
For each mode we obtain
\begin{equation}
L_j A_j = A_{j, 0} \exp[-i\mathbf{k}^{(j)}\cdot \mathbf{u}(\mathbf{r})] \left\{\cdots\right\}
\end{equation}
with
\begin{eqnarray}
\left\{\cdots\right\} &=& -k_\beta^{(j)} k_\gamma^{(j)} (\partial_\alpha u_\beta) (\partial_\alpha u_\gamma) + 2 k_\alpha^{(j)} k_\beta^{(j)} \partial_\alpha u_\beta \nonumber \\
&& - i k_\beta^{(j)} \partial_\alpha^2 u_\beta.
\end{eqnarray}
In terms of the tensor (\ref{strangeEulerStrain}) we can rewrite this as 
\begin{equation}
\left\{\cdots\right\} = 2 k_\alpha^{(j)} k_\beta^{(j)} \bar{e}_{\alpha\beta} - i k_\beta^{(j)} \partial_\alpha^2 u_\beta.
\end{equation}
Therefore the elastic energy density for fixed amplitudes is for each mode $j$
\begin{eqnarray}
f_{nonlocal}^{(j)} &=& |A_{j,0}|^2 |\left\{\cdots\right\}|^2 \nonumber\\
&=& 4 |A_{j,0}|^2 \left(k_\alpha^{(j)} k_\beta^{(j)} \bar{e}_{\alpha\beta}\right)^2   \nonumber \\
&& + |A_{j,0}|^2 \left( k_\beta^{(j)} \partial_\alpha^2 u_\beta \right)^2.
\end{eqnarray}
The first term corresponds to an elastic term, the second to a strain gradient contribution.
For long wave distortions the second term is negligible and will not be considered here.
Whereas the energy expression for the individual modes contains the strain tensor $\bar{e}_{ij}$, which may be somewhat unexpected from point of view of elasticity, the situation changes if we sum over the three modes.
We then get for the elastic term
\begin{equation} \label{intermediateresult}
f_{nonlocal} = 3 |A_0|^2 \bar{\Delta},
\end{equation}
with
\begin{equation} \label{eq33}
\bar{\Delta} = \frac{3}{2} \bar{e}_{xx}^2 + \frac{3}{2} \bar{e}_{yy}^2 + 2 \bar{e}_{xy}^2 + \bar{e}_{xx}\bar{e}_{yy}.
\end{equation}
Here we skipped the strain gradient term and assumed that all amplitudes have the same magnitude $|A_{j, 0}| = |A_0|$.
This expression is analogous to the one in Refs.~\onlinecite{Chan:2009aa, Huter:2015aa}, which contain an incorrect sign in the definition of the displacement.
Therefore, here the Eulerian variant of the strain tensor appears.
This correction is important as it reflects that materials get stiffer (softer) under compression (tension), and not vice versa.
For the present case of triangular systems, this expression coincides with the one defined through the Euler-Almansi strain,
\begin{equation} \label{eq34}
{\Delta} = \frac{3}{2} {e}_{xx}^2 + \frac{3}{2} {e}_{yy}^2 + 2 {e}_{xy}^2 + {e}_{xx}{e}_{yy},
\end{equation}
hence $\Delta=\bar{\Delta}$ and we explain this coincidence in section \ref{sectionBCC} in detail.
This implies again that for constant amplitudes the non-linear elasticity is described entirely through geometric non-linearity.

\begin{figure}
\begin{center}
\includegraphics[width=8.5cm]{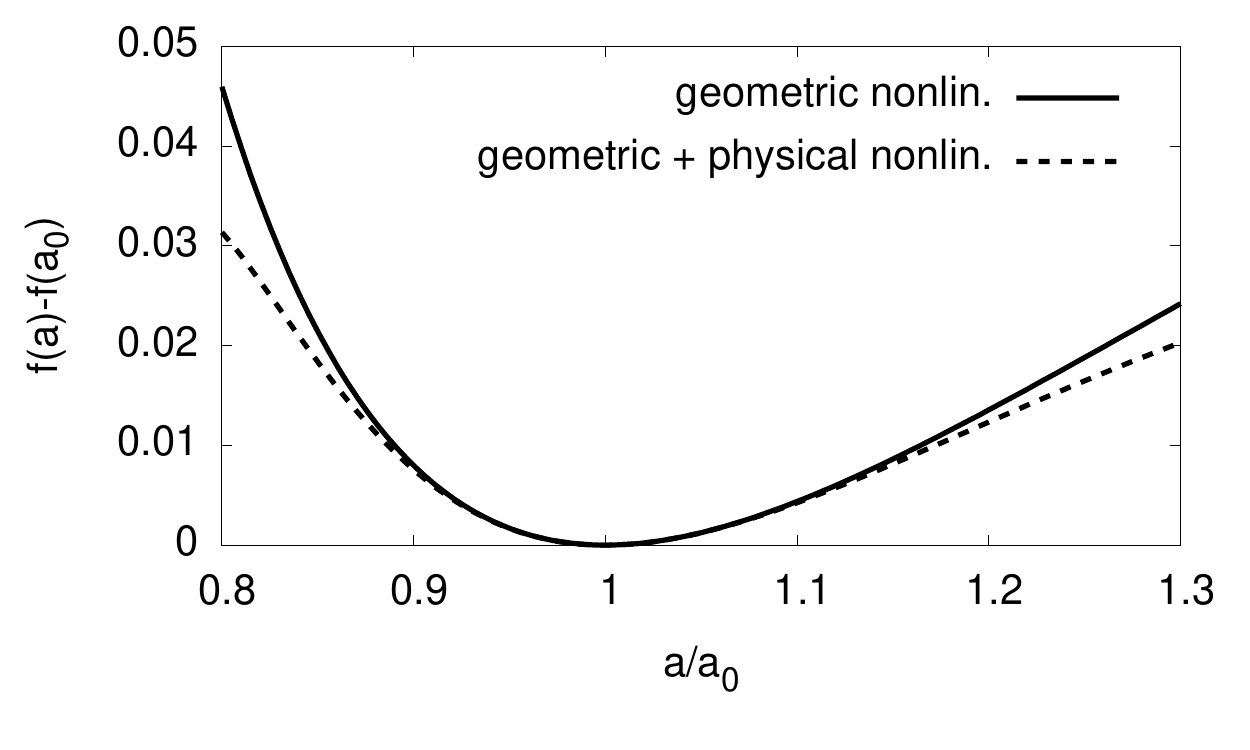}
\caption{
Elastic energy density in the two-dimensional triangular PFC model, as a function of the relative lattice constant $a/a_0$ for isotropic straining.
The parameters $\bar{\psi}=0.3$ and $\pfcepsilon=0.6$ are used. 
The solid curve is for fixed amplitude $A=A(\Delta=0)$, whereas the dashed curve includes the physical non-linearity due to the elastic weakening $A=A(\Delta)$.
}
\label{fig3}
\end{center}
\end{figure}

In Fig.~\ref{fig3} we show the elastic energy for the particular case of isotropic straining as a function of the lattice constant, $a/a_0$, where the geometric non-linearity results in the material indeed becoming stiffer (softer) under compression (tension).
The Eulerian strains are
\begin{equation}
e_{xx}=e_{yy}=\frac{a-a_0}{a} - \frac{1}{2}\left( \frac{a-a_0}{a} \right)^2,\qquad e_{xy}=0.
\end{equation}

If we now take the situation of isotropic deformations and also minimise the energy with respect to the amplitudes, all of them change their magnitude equally due to symmetry.
Hence we have $A_j = A_{j,0}\exp(-i\mathbf{k}^{(j)}\cdot\mathbf{u})$ with the same real and positive value $A_0=-A_{1,0}=A_{2,0}=A_{3,0}$.
Evaluation of the free energy density $f=f_{local}+f_{nonlocal}$ and minimisation with respect to $A_0$ gives
\begin{equation} \label{isotropicweakening2D}
A_0(\Delta) = \frac{1}{5} \left( \bar{\psi} + \frac{1}{3} \sqrt{9\bar{\psi}^2 + 15(\Gamma-\Delta)}\right),
\end{equation}
which is the same as in Ref.~\onlinecite{Chan:2009aa}, written here for the case $\bar{\psi}>0$.
We note that for large values of $\pfcepsilon\gg \Delta$ the amplitudes hardly change with the strain, and then geometric non-linearity is essentially the only source for deviations from linear elasticity, as before in the one-dimensional case.
The energy density is (again for general values of $\pfcepsilon$ and $\Delta$)
\begin{equation}
f(\Delta) = \frac{45}{2}A_0^4(\Delta) - 12\bar{\psi}A_0^3(\Delta) - 3(\Gamma-\Delta)A_0^2(\Delta),
\end{equation}
which is valid for isotropic deformations.
Again the strain dependent amplitudes lead to physical non-linearity.
Due to the amplitude as additional degree of freedom, which is used here for minimization, the energy is lower than for fixed amplitude, see Fig.~\ref{fig3}.

The assumption of all amplitudes being the same in magnitude is valid for isotropic deformations only.
Although the expression involving $\Delta$ may suggest that it holds also for other cases, this is not the case.
For anisotropic deformations the amplitudes will in general change differently as a function of the applied strain.

\begin{figure}
\begin{center}
\includegraphics[width=8.5cm]{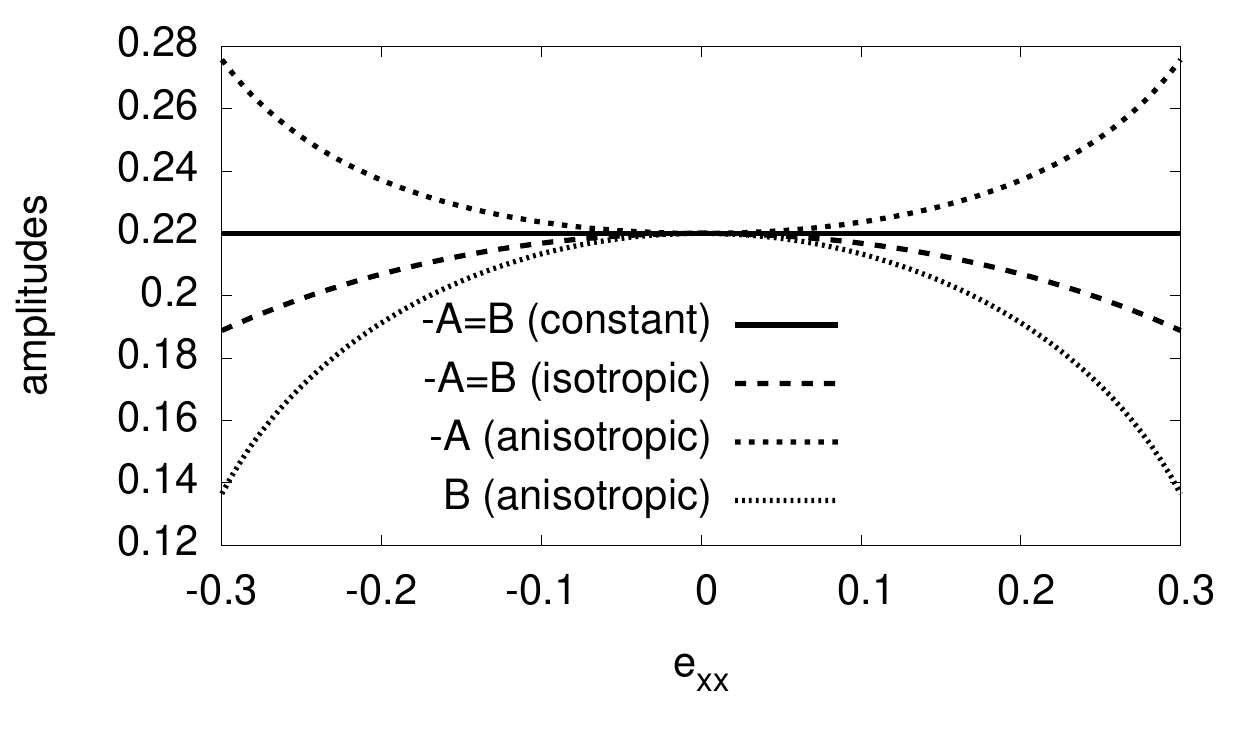}
\caption{Absolute values of the amplitudes as a function of the uniaxial strain $e_{xx} = \bar{e}_{xx}$ for $e_{xy}=e_{yy}=0$ in the two-dimensional PFC model.
The solid line is the case of strain independent amplitudes, where only geometric non-linearity arises.
The long dashed curve uses the approximation of equal strain dependence of the amplitudes according to Eq.~(\ref{isotropicweakening2D}), where all amplitudes are subject to the same elastic weakening.
For that, the expression (\ref{eq33}) is used, with $\bar{e}_{xx}$ being the only nonvanishing component.
The remaining two curves show the unequal response of the amplitudes as a result of the uniaxial strain.
Parameters are $\pfcepsilon=0.6$, $\bar{\psi}=0.3$.}
\label{fig5}
\end{center}
\end{figure}

For general amplitudes $A_j = A_{j,0}\exp(-i\mathbf{k}^{(j)}\cdot\mathbf{u})$ the nonlocal energy contribution becomes
\begin{eqnarray}
f_{nonlocal} &=& 4 |A_{1,0}|^2 \bar{e}_{yy}^2 \nonumber \\
&+& 4  |A_{2,0}|^2 \left ( \frac{3}{4} \bar{e}_{xx} - \frac{\sqrt{3}}{2} \bar{e}_{xy} + \frac{1}{4} \bar{e}_{yy} \right)^2 \nonumber \\
&+& 4  |A_{3,0}|^2 \left ( \frac{3}{4} \bar{e}_{xx} + \frac{\sqrt{3}}{2} \bar{e}_{xy} + \frac{1}{4} \bar{e}_{yy} \right)^2.
\end{eqnarray}
From now on we assume that all prefactors $A_{j,0}$ are real.
Then the local energy density reads
\begin{eqnarray}
f_{local} &=& -\Gamma (A_{1,0}^2 + A_{2,0}^2 + A_{3,0}^2) \nonumber \\
&+& \frac{3}{2} \Big( A_{1,0}^4 + A_{2,0}^4 + A_{3,0}^4 + 4A_{1,0}^2A_{2,0}^2+ 4A_{1,0}^2A_{3,0}^2 \nonumber \\
&+& 4A_{2,0}^2A_{3,0}^2 \Big) + 12\bar{\psi} A_{1,0} A_{2,0}A_{3,0}.
\end{eqnarray}
We have to minimise (for given strain) the energy with respect to all amplitudes $A_{j,0}$.
To simplify the situation, we consider the case of uniaxial stretching in $x$ direction, i.e.~$\bar{e}_{xy}=\bar{e}_{yy}=0$.
Then by symmetry two amplitudes are equal, and we write $A_{1,0}=A<0$ and $A_{2,0}=A_{3,0}=B>0$.
With this the energy densities become
\begin{eqnarray}
f_{local} &=& -\Gamma(A^2+2B^2) + \frac{3}{2} (A^4+6B^4+8A^2B^2) \nonumber \\
&&+ 12AB^2\bar{\psi}
\end{eqnarray}
and
\begin{equation}
f_{nonlocal} = \frac{9}{2} B^2 \bar{e}_{xx}^2.
\end{equation}
Minimization of $f$ has to be performed with respect to $A$ and $B$.
From the minimisation with respect to $A$ we get
\begin{equation}
B = \sqrt{\frac{(\Gamma-3A^2)A}{6(2A+\bar{\psi)}}},
\end{equation}
where we have chosen the branch $B>0$.
From the minimisation with respect to $B$ we get the condition
\begin{equation}
-4\Gamma+ 36 B^2 +24A^2 + 24 A\bar{\psi} + 9\bar{e}_{xx}^2=0.
\end{equation}
Fig.~\ref{fig5} shows the amplitudes as a function of the applied strain.
As one can see the amplitudes indeed depend differently on the strain.
The mode related to $\mathbf{k}^{(1)}$, which has a RLV perpendicular to the applied load, increases in magnitude as a function of strain;
this mode does not carry elastic energy, therefore its increase in magnitude is not penalised.
In contrast, the other two modes decrease in magnitude, and this more strongly than in the isotropic approximation.

\begin{figure}
\begin{center}
\includegraphics[width=8.5cm]{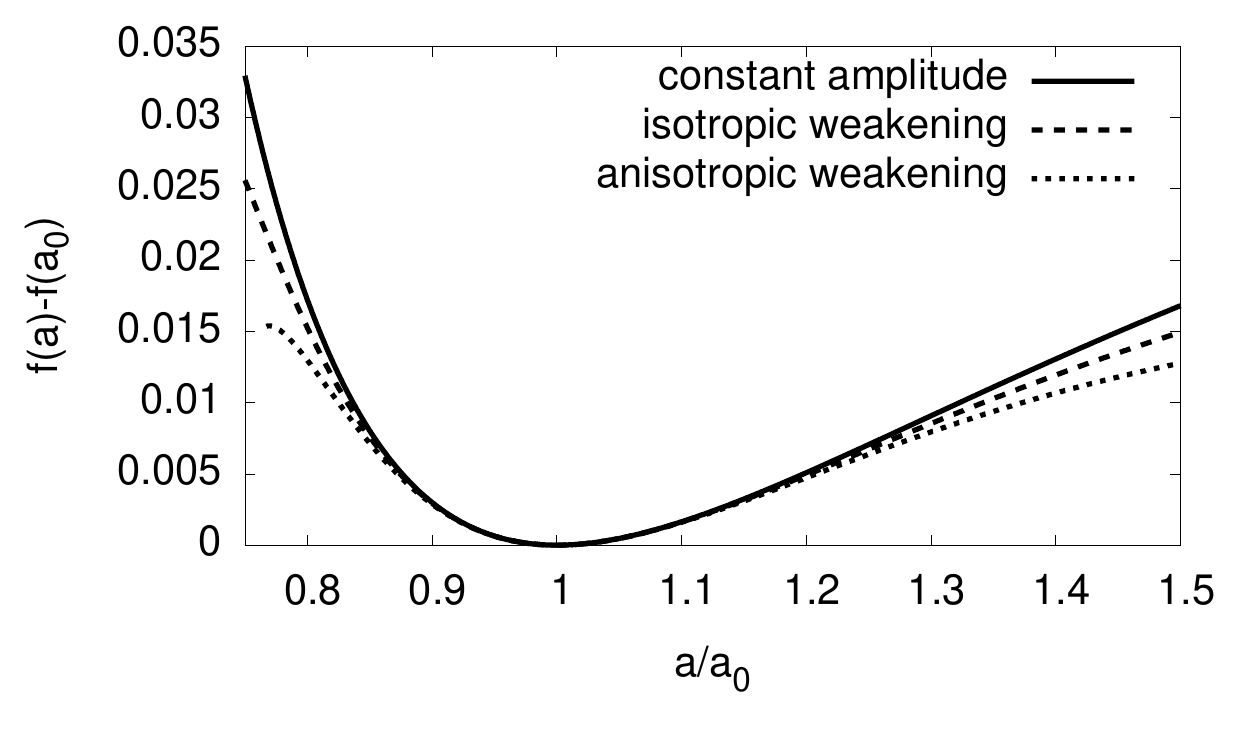}
\caption{Elastic energy density of the two-dimensional PFC model for a uniaxial strain $e_{xx} = \bar{e}_{xx}$ and $e_{xy}=e_{yy}=0$.
The elastic energy is highest in the non-linear regime if the amplitudes are considered as constant (solid curve).
The equal dependence on the strain, $-A_{1,0}=A_{2,0}=A_{3,0}$ reduces the energy and leads to the dashed curve.
The dotted curve correctly considers unequal weakening of the amplitudes due to strain and leads to the lowest elastic energy.
Parameters are $\pfcepsilon=0.6$, $\bar{\psi}=0.3$.
}
\label{fig5a}
\end{center}
\end{figure}

The energy density can then be written as
\begin{equation}
f = f - \frac{1}{2} \frac{\partial f}{\partial B} B = -\Gamma A^2 + \frac{3}{2} A^4 - 9B^4,
\end{equation}
where the partial derivative is zero by the minimisation condition and leads to the first identity.
The total energy as a function of the uniaxial change of the lattice constant is shown in Fig.~\ref{fig5a}.
As expected, the energy is lower in the full anisotropic description compared to the isotropic approximation.
The reason is that we allow for additional degrees of freedom, $A\neq-B$, which allow to further reduce the energy.
By this, the contribution of physical non-linearity to the elastic response becomes more important.

\section{Body-centered cubic materials}
\label{sectionBCC}

The body-centred cubic (bcc) phase exists in equilibrium in some parameter regions of the three-dimensional phase field crystal model.
Again we use a one-mode approximation according to Eq.~(\ref{onemode}), this time summing over $N=6$ normalised reciprocal lattice vectors,
\begin{eqnarray}
&&\mathbf{k}_{110}= \left(
\begin{array}{c} 
1/\sqrt{2}\\
1/\sqrt{2}\\
0
\end{array}
\right),
\qquad 
\mathbf{k}_{101} = \left(
\begin{array}{c} 
1/\sqrt{2}\\
0\\
1/\sqrt{2}
\end{array}
\right), 
\nonumber \\
&&
\mathbf{k}_{011} = \left(
\begin{array}{c} 
0\\
1/\sqrt{2}\\
1/\sqrt{2}
\end{array}
\right),
\qquad
\mathbf{k}_{1\bar{1}0} = \left(
\begin{array}{c} 
1/\sqrt{2}\\
-1/\sqrt{2}\\
0
\end{array}
\right),
\nonumber \\ 
&&\mathbf{k}_{10\bar{1}} = \left(
\begin{array}{c} 
1/\sqrt{2}\\
0\\
-1/\sqrt{2}
\end{array}
\right),
\qquad
\mathbf{k}_{01\bar{1}} = \left(
\begin{array}{c} 
0\\
1/\sqrt{2}\\
-1/\sqrt{2}
\end{array}
\right). \label{3DRLVs}
\end{eqnarray}

Similarly to above we get by insertion into the free energy and orthogonality (see Appendix \ref{appendix::aederivation} and Refs.~\onlinecite{Wu:2007kx, Spatschek:2010fk})
\begin{eqnarray}
F &=& \int d\mathbf{r} \Bigg[ 4 \sum_{j=1}^6 \left|\Box_j A_j\right|^2 + (3\bar{\psi}^2-\pfcepsilon) \sum_{j=1}^{6} A_j A_j^* \nonumber \\
&&+ 3 \Bigg\{ \left( \sum_{j=1}^{6} A_j A_j^* \right)^2  - \frac{1}{2}\sum_{j=1}^{6} |A_j|^4 \nonumber \\
&&+ 2A_{110}^* A_{1\bar{1}0}^* A_{101} A_{10\bar{1}} + 2A_{110} A_{1\bar{1}0} A_{101}^* A_{10\bar{1}}^* 
\nonumber \\
&& + 2A_{1\bar{1}0} A_{011} A_{01\bar{1}} A_{110}^* 
+ 2A_{1\bar{1}0}^* A_{011}^* A_{01\bar{1}}^* A_{110} 
\nonumber \\
&& + 2A_{01\bar{1}} A_{10\bar{1}}^* A_{101} A_{011}^*
+ 2A_{01\bar{1}}^* A_{10\bar{1}} A_{101}^* A_{011} \Bigg\} 
\nonumber \\
&&
+6\bar{\psi}  \Big( A_{011}^* A_{101} A_{1\bar{1}0}^* + A_{011} A_{101}^* A_{1\bar{1}0} + A_{011}^* A_{110} A_{10\bar{1}}^* \nonumber \\
&&
+ A_{011} A_{110}^* A_{10\bar{1}} 
+ A_{01\bar{1}}^* A_{110} A_{101}^* + A_{01\bar{1}} A_{110}^* A_{101} 
\nonumber \\
&& 
+ A_{01\bar{1}}^* A_{10\bar{1}} A_{1\bar{1}0}^* + A_{01\bar{1}} A_{10\bar{1}}^* A_{1\bar{1}0} \Big) \nonumber \\
&&
+ \frac{1}{2} \bar{\psi}^2(1-\pfcepsilon) + \frac{1}{4} \bar{\psi}^4 \Bigg],  \label{nonlinAE}
\end{eqnarray}
expressed here through the box operator 
\begin{equation}
\Box_j=\mathbf{k}_j\cdot\nabla - \frac{i}{2q_0}\nabla^2 =-\frac{i}{2q_0} L_j
\end{equation}
with $q_0=|\mathbf{k}_j|=1$ and $\bar{\psi}<0$.

\begin{figure}
\begin{center}
\includegraphics[width=8.5cm]{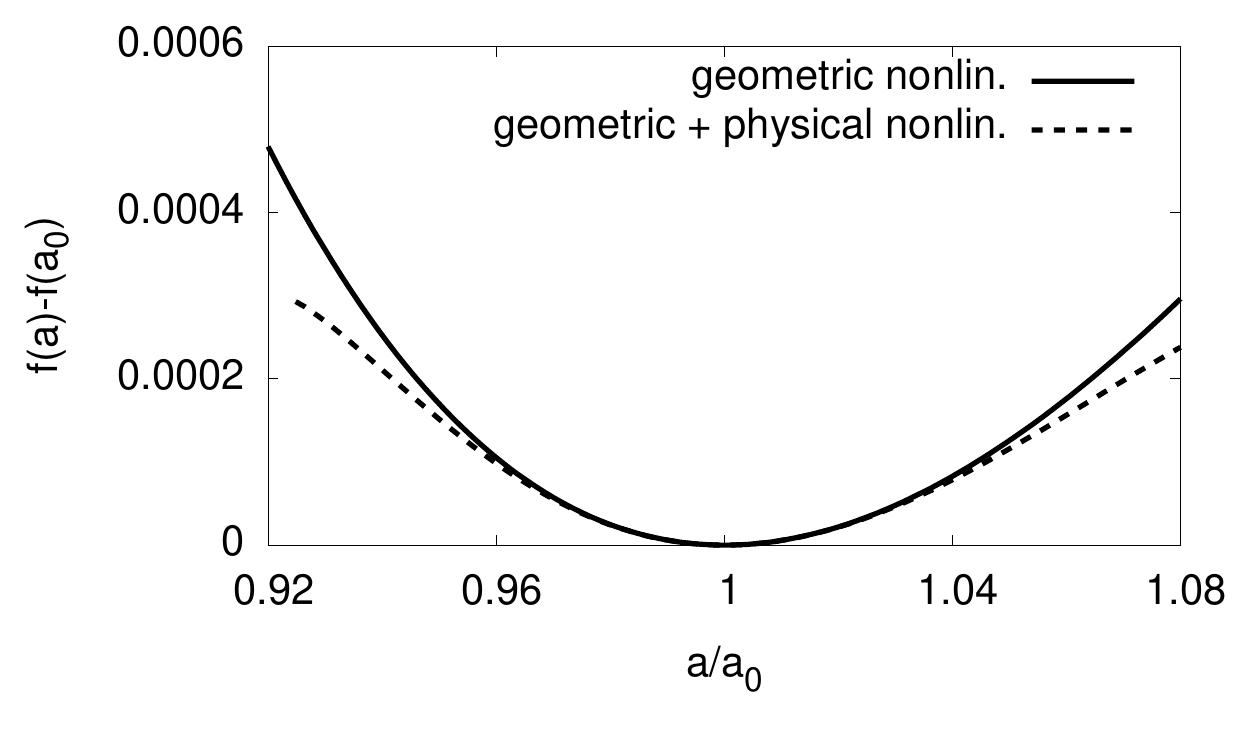}
\caption{
Elastic energy density in the three-dimensional bcc PFC model, as a function of the relative lattice constant $a/a_0$ for isotropic straining.
The parameters $\bar{\psi}=-0.18$ and $\pfcepsilon=0.1$ are used. 
The solid curve is for fixed amplitude $A_0=A(\Delta=0)$, whereas the dashed curve includes the physical non-linearity due to the elastic weakening $A_0=A_0(\Delta)$, see Eq.~(\ref{isotropicamplitudebcc}).
}
\label{figaa}
\end{center}
\end{figure}

The nonlocal contribution from the box operator can be evaluated as before, and we get
\begin{equation}
F_{nonlocal} = 4 \int d{\mathbf r} \sum_{j=1}^6 \left|\Box_j A_j\right|^2 = 4 \int d\mathbf{r} \bar{\Delta} |A_0|^2
\end{equation}
with
\begin{eqnarray}
\bar{\Delta} &=& \bar{e}_{xx}^2 + \bar{e}_{yy}^2 + \bar{e}_{zz}^2 
+   2 (\bar{e}_{xy}^2 + \bar{e}_{yz}^2 + \bar{e}_{xz}^2) \nonumber \\
&& + \bar{e}_{xx}\bar{e}_{yy}+ \bar{e}_{yy} \bar{e}_{zz} + \bar{e}_{xx} \bar{e}_{zz}, \label{eq49}
\end{eqnarray}
where strain gradient terms are suppressed.
Here we have assumed that all amplitudes have the same magnitude, $A_{j, 0}=A_0$.
This is the case for an isotropic deformation $\bar{e}_{ij} = \bar{e}\delta_{ij}$, which leads to
\begin{eqnarray}
F &=& \int d\mathbf{r} \Big\{ \left[4\bar{\Delta}+6(3\bar{\psi}^2-\pfcepsilon)\right]  A_0^2 + 48 \bar{\psi} A_0^3 \nonumber \\
&&+ 135 A_0^4 + \frac{1}{2} \bar{\psi}^2(1-\pfcepsilon) + \frac{1}{4} \bar{\psi}^4 \Big\}. \label{eq50}
\end{eqnarray}
For fixed amplitudes $A_0$ we therefore see that --- as before for the one- and two-dimensional case --- the elastic part of the energy is linear in $\bar{\Delta}$ and therefore quadratic in the (Eulerian) strains $\bar{e}_{ij}$.
As before, this gives rise to the geometric non-linearity, see Fig.~\ref{figaa}.

We point out that here only the strain tensor $\bar{e}_{ij}$ as defined in Eq.~(\ref{strangeEulerStrain}) allows the compact notation of the elastic energy through $\bar{\Delta}$, similar to the two-dimensional case.
In contrast, it is here not possible to represent the elastic energy in terms of $\Delta$ directly, which is defined via the Euler strains $e_{ij}$.
The reason is that in the two dimensional triangular case (indicated here through a superscript `tri') both $\Delta^{tri}$ and $\bar{\Delta}^{tri}$ can be expressed through the (identical) traces of the tensors $e$ and $\bar{e}$ or powers of them.
Explicitly, one gets from Eqs.~(\ref{eq33}) and (\ref{eq34})
\begin{eqnarray}
\Delta^{tri} &=& \mathrm{tr}(e^2) + \frac{1}{2} \mathrm{tr}(e)^2, \\
\bar{\Delta}^{tri} &=& \mathrm{tr}(\bar{e}^2) + \frac{1}{2} \mathrm{tr}(\bar{e})^2, 
\end{eqnarray}
indicating elastic isotropy \cite{LandauLifshitz:7}.
In contrast, for the bcc system we cannot expect the equality of $\bar{\Delta}$ and $\Delta$ due to the cubic symmetry. 
Indeed, such a representation is not possible for the three-dimensional bcc expression (\ref{eq49}) and an analogous term $\Delta$ involving $e_{ij}$.
One can readily check that the equivalence of $\bar{\Delta}$ and $\Delta$ fails for specific situations with nonvanishing shear.

For an isotropic deformation $e_{ij}=e_{xx}\delta_{ij}$ we can identify in the small strain regime (where all strain tensors coincide and the deformed and reference volume are the same) the bulk modulus $K$ by the comparison with the elastic part of the energy $F_{nonlocal}=9 K V e_{xx}^2/2$ as $K=16 |A_0|^2/3$.

In the following we relax the amplitudes to obtain physical non-linearity, first again for the isotropic and then an anisotropic situation.

If we minimise the energy (\ref{eq50}) with respect to $A_0$ in the isotropic case, it becomes a function of $\bar{\Delta}$. 
Explicitly, we get
\begin{equation} \label{isotropicamplitudebcc}
A_0(\bar{\Delta}) = \frac{1}{45} \left(-6\bar{\psi} + \sqrt{3} \sqrt{15\pfcepsilon-33\bar{\psi}^2-10\bar{\Delta}}\right).
\end{equation}
In this case the free energy becomes a non-linear function of $\bar{\Delta}$, see Fig.~\ref{figaa}.
This is important, as through the quadratic nature of $\bar{\Delta}$ the elastic energy is symmetric if plotted versus the isotropic strain $\bar{e}_{xx}=\bar{e}_{yy}$, even with strain dependent amplitudes.
Already at this point we mention that this outcome suggests to inspect the elastic energy of real materials as a function of $\bar{e}_{ij}$ instead of the Lagrangian variant $\epsilon_{ij}$.
This will be pursued in the following section.

\begin{figure}
\begin{center}
\includegraphics[width=8.5cm]{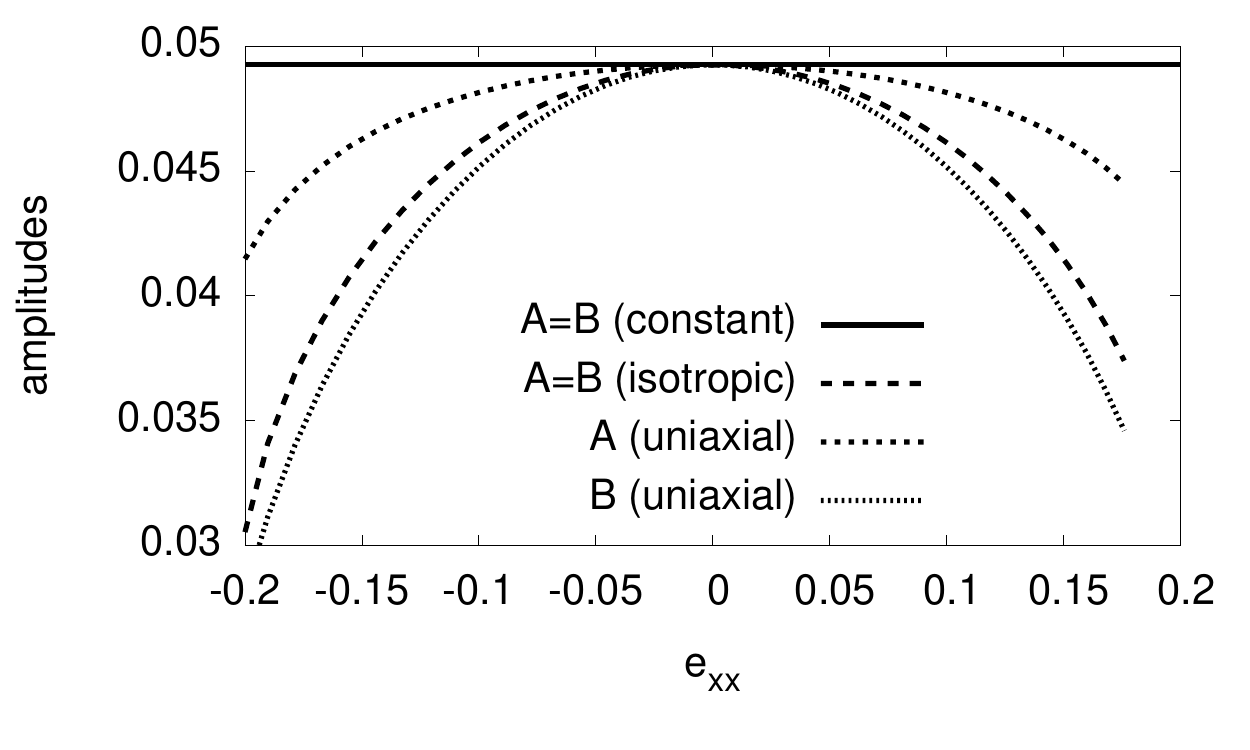}
\caption{Values of the amplitudes for the bcc model as a function of the uniaxial strain $e_{xx} = \bar{e}_{xx}$, and all other strain components vanish, $\bar{e}_{ij}=0$.
The solid line is the case of strain independent amplitudes.
The dashed curve is based on the isotropic approximation, where all amplitudes equally depend on the strain according to Eqs.~(\ref{eq49}) and (\ref{isotropicamplitudebcc}).
The two dotted curves are based on the true minimization with two independent amplitudes $A$ and $B$.
Parameters are $\pfcepsilon=0.1$, $\bar{\psi}=-0.18$.
}
\label{fig5d1}
\end{center}
\end{figure}

\begin{figure}
\begin{center}
\includegraphics[width=8.5cm]{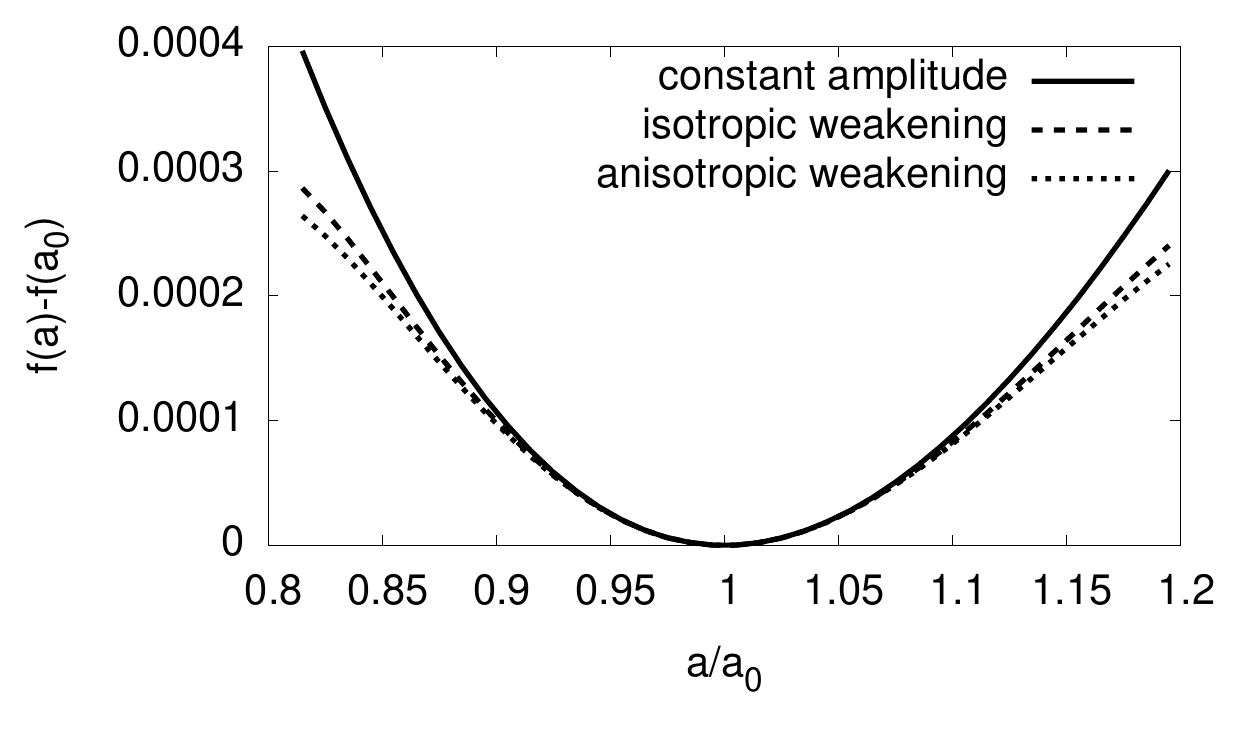}
\caption{
Elastic energy density of the three-dimensional bcc model for a uniaxial strain $e_{xx} = \bar{e}_{xx}$, and all other strain components vanish.
The elastic energy is highest in the non-linear regime if the amplitudes are considered as constant (solid curve).
The equal dependence of the amplitudes on the strain according to Eqs.~(\ref{eq49}) and (\ref{isotropicamplitudebcc}) reduces the energy and leads to the dashed curve.
Consideration of anisotropic weakening leads to the lowest energy (dotted curve).
Parameters are $\pfcepsilon=0.1$, $\bar{\psi}=-0.18$.
}
\label{fig5d2}
\end{center}
\end{figure}

We conclude the analysis similar to the previous two-dimensional case with the situation of an anisotropic strain, where the amplitudes depend differently on the mechanical load.
We use a homogeneous uniaxial strain $e_{xx}=\bar{e}_{xx}$, assuming that all other strain components $\bar{e}_{ij}$ vanish.
In this case the amplitudes group into two sets, having the same magnitude in each of these groups.
The first set contains amplitudes with RLVs perpendicular to the strain, i.e.~$\mathbf{k}_j\cdot\hat{\mathbf{x}}=0$, namely $A:=A_{011, 0}=A_{01\bar{1}, 0}$.
The remaining amplitudes with $\mathbf{k}_j\cdot\hat{\mathbf{x}}\neq 0$ have the same magnitude, denoted as $B$, i.e.~$B:=A_{110, 0}=A_{1\bar{1}0, 0}=A_{101, 0}=A_{10\bar{1}, 0}$.
With $\bar{\Delta} = \bar{e}_{xx}^2$ we obtain the free energy density
\begin{eqnarray}
&&f = 9 A^4 + 48 AB^2 \bar{\psi} + A^2 (72 B^2 - 2 \pfcepsilon + 6 \bar{\psi}^2) \\ \nonumber 
&& + \frac{1}{4} \left[ 216 B^4 + \bar{\psi}^2 (2-2\pfcepsilon + \bar{\psi}^2) + 16 B^2 (\bar{\Delta} - \pfcepsilon + 3 \bar{\psi}^2) \right]. 
\end{eqnarray}
By minimization with respect to $A$ and $B$ we obtain the strain dependent amplitudes.
The isotropic and anisotropic amplitude relaxation is shown in Fig.~\ref{fig5d1} and its effect on the free energy in Fig.~\ref{fig5d2}.
For the used parameters the influence of the amplitude relaxation (physical non-linearity) is weaker than for the previous one- and two-dimensional cases, where we used a lower undercooling $\pfcepsilon$.
For smaller values of $\pfcepsilon$ we find here a rather small range of strains before the phase becomes unstable.
This is related to the narrow single phase region of bcc in the phase diagram.
Also, the influence of anisotropic versus isotropic amplitude relaxation is lower than for the two-dimensional triangular model.
However, we recall that already a decrease of few percent in the free energy can significantly influence phase coexistence regimes. 
In the low temperature limit the non-linear elastic energy stems from geometric non-linearity alone, and then the elastic energy becomes a quadratic function of the strain components $\bar{e}_{ij}$.
This prediction will be compared to {\em ab initio} results in the following section.

\section{The Birch-Murnaghan equation: Comparison with ab initio simulations}
\label{BirchMurnaghanSection}

The Murnaghan\cite{MurnaghanFD:1944pnas} and Birch-Murnaghan equations of state\cite{BirchF:1947pr} are used to describe the non-linear elastic response under isotropic stretching or compression.
They are frequently used to fit {\em ab initio} data for energy-volume curves.
Birch\cite{BirchF:1947pr} emphasises the importance of the distinction between Lagrangian and Eulerian descriptions, and notes that the representation is simpler in the Eulerian frame.
Our findings support this concept from a PFC perspective.
It is therefore the goal of this section to compare the PFC model with the classical equations for energy-volume curves and to further link it to {\em ab initio} calculated energy-strain curves of elementary bcc systems and graphene.

The Birch-Murnaghan model describes the energy as a function of volume as
\begin{eqnarray} \label{BM3D}
E_\mathrm{BM}(V) &=& E_0 + \frac{9 V_0 K}{16} \Bigg\{ \left[ \left( \frac{V_0}{V} \right)^{2/3} -1 \right]^3 K' \nonumber \\
&+&  \left[ \left( \frac{V_0}{V} \right)^{2/3} -1 \right]^2 \left[ 6-4\left( \frac{V_0}{V}\right)^{2/3} \right] \Bigg\}.
\end{eqnarray}
Here $V_0$ is the equilibrium volume, $V$ the actual volume of the isotropically deformed system, $K$ the zero pressure bulk modulus and $K'=(dK/dP)_{P=0}$ the derivative of the bulk modulus.
The latter quantity is (usually) positive, as materials get stiffer under compression.
The above equation is applicable for three dimensions.

One can derive the pressure for the present isotropic case from the standard thermodynamic relation
\begin{equation} \label{thermointen1}
P = -\left( \frac{\partial E_\mathrm{BM}}{\partial V} \right)_{N, T}
\end{equation}
and from this get the bulk modulus as 
\begin{equation} \label{thermointen2}
K(V) = - V \left( \frac{\partial P}{\partial V} \right)_{N, T}.
\end{equation}
In the limit of vanishing pressure, i.e.~for $V\to V_0$, one gets the leading contribution, which is denoted above as a constant $K$.
The derivative of the bulk modulus for zero pressure then follows from
\begin{equation} \label{thermointen3}
K'= \lim_{V\to V_0} \frac{\displaystyle \left( \frac{\partial K}{\partial V}\right)_{N, T}}{\displaystyle \left( \frac{\partial P}{\partial V}\right)_{N, T}}.
\end{equation}

In order to link this equation to the phase field crystal model we rewrite the Birch-Murnaghan expression (\ref{BM3D}) in terms of the Eulerian strain,
\begin{equation}
e_{xx} = e_{yy} = e_{zz} = \frac{a-a_0}{a} - \frac{1}{2} \left( \frac{a-a_0}{a} \right)^2 = \frac{a^2-a_0^2}{2a^2}.
\end{equation}
Here we particularly have $\bar{e}_{xx}=e_{xx}$, $\bar{e}_{yy}=e_{yy}$, $\bar{e}_{zz}=e_{zz}$. 
With $V=a^3$ and $V_0=a_0^3$ we therefore get in three dimensions
\begin{equation} \label{Mur3}
\left( \frac{V_0}{V} \right)^{2/3} = 1-2e_{xx}
\end{equation}
and arrive at the compact representation
\begin{equation} \label{Mur1}
E_\mathrm{BM}(e_{xx}) = \frac{9}{2} K V_0 e_{xx}^2 \left[ 1+ (4-K') e_{xx} \right].
\end{equation}
For small strains $|e_{xx}|\ll 1$ it reduces to the usual linear elastic energy $E_\mathrm{BM}(V) \approx 9 KV_0 \epsilon_{xx}^2/2$.
For many materials the bulk modulus derivative turns out to be close to $K'=4$, and in this case we obtain the simple expression
\begin{equation} \label{Mur2}
E_{\mathrm{BM}, K'=4}(e_{xx}) = \frac{9}{2} K V_0 e_{xx}^2,
\end{equation}
which also holds in the non-linear regime.
Notice that the reference volume $V_0$ instead of the actual volume $V$ appears here.
As will be shown below this formula fits very well the {\em ab initio} data for various elemental metals.
From  Eq.~(\ref{Mur1}) we see that deviations from $K'=4$ break the symmetry between compression and expansion $e_{xx}\to -e_{xx}$.

The older Murnaghan model\cite{MurnaghanFD:1944pnas} is given by
\begin{eqnarray}
E_\mathrm{M}(V) &=& E_0 + K V_0 \Bigg[ \frac{1}{K'(K'-1)} \left( \frac{V}{V_0} \right)^{1-K'} \nonumber \\
&+&  \frac{1}{K'} \frac{V}{V_0} - \frac{1}{K'-1} \Bigg].
\end{eqnarray}
Expanding it in terms of the Euler strain gives
\begin{eqnarray}
E_\mathrm{M}(e_{xx}) &=& \frac{9}{2} K V_0 e_{xx}^2 \Bigg[ 1+ (4-K') e_{xx} \nonumber \\
&+& \frac{1}{12}(143-63K'+9{K'}^2) e_{xx}^2 + {\cal O}(e_{xx}^3) \Bigg].
\end{eqnarray}
It agrees with the Birch-Murnaghan model up to third order in $e_{xx}$.

In comparison the three-dimensional bcc phase field crystal model with constant amplitudes delivers the comparable expression for the averaged elastic free energy density
\begin{equation} \label{eq65}
f_\mathrm{PFC}(e_{xx}) = \frac{9}{2} K e_{xx}^2,
\end{equation}
with the identification $K = 16|A_0|^2/3$.
In the spirit of the discussion in Section \ref{pfcintrosec} we can therefore conclude that the phase field crystal is analogous to the Birch-Murnaghan model for $K'=4$ in the low temperature limit.
This includes in particular that the elastic energy is symmetric with respect to the Eulerian strain for bulk deformations.
As discussed before, the effect of strain dependent amplitudes for isotropic deformations can lower the elastic energy, which can lead to deviations from the Birch-Murnaghan curve.
Still, the PFC energy will remain quadratic in the strains, and therefore in particular symmetric under the exchange $e_{xx}\to-e_{xx}$.
Hence it affects only higher order corrections of the elastic energy starting at ${\cal O}(e_{xx}^4)$.
This effect is most pronounced for small values of $\epsilon$, which corresponds to high temperatures.

\begin{figure}
\begin{center}
\includegraphics[width=8.5cm]{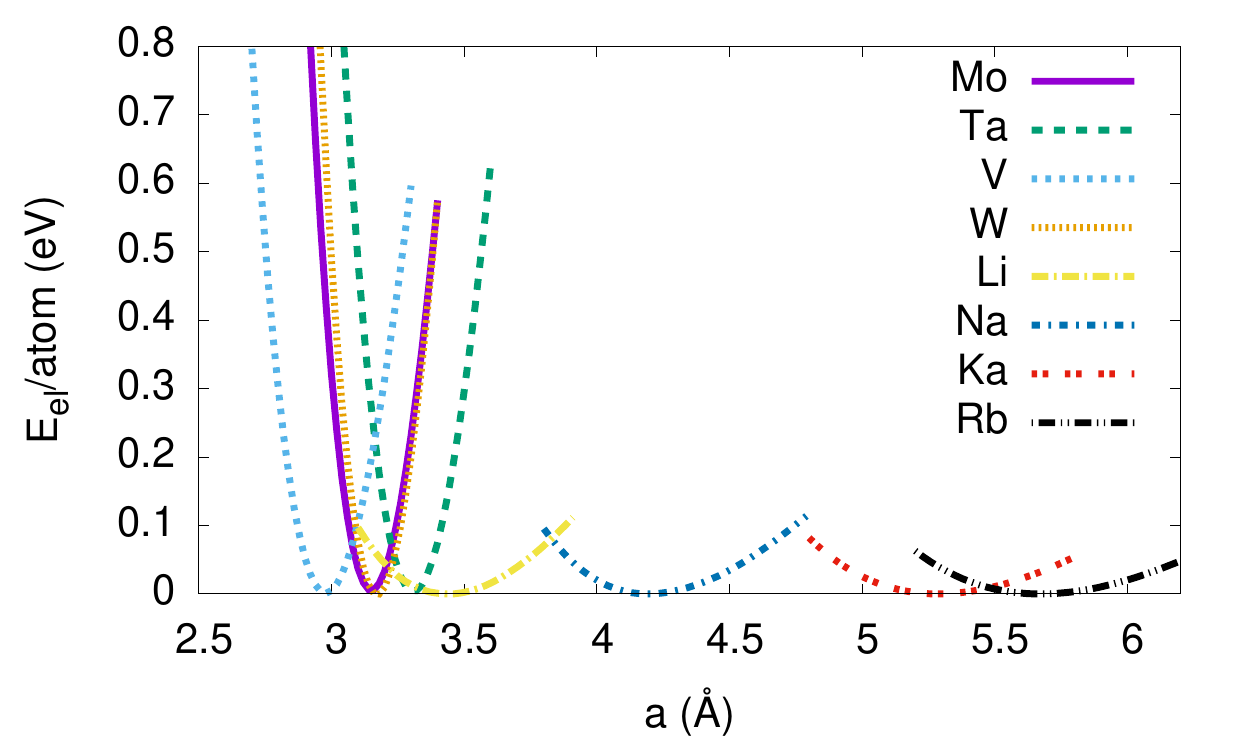}
\caption{(Color online) Elastic energy as a function of the lattice constant for various nonmagnetic bcc metals. Notice that in this representation the energy is not symmetric around the minimum position $a_0$.}
\label{figabinitio1}
\end{center}
\end{figure}
\begin{figure}
\begin{center}
\includegraphics[width=8.5cm]{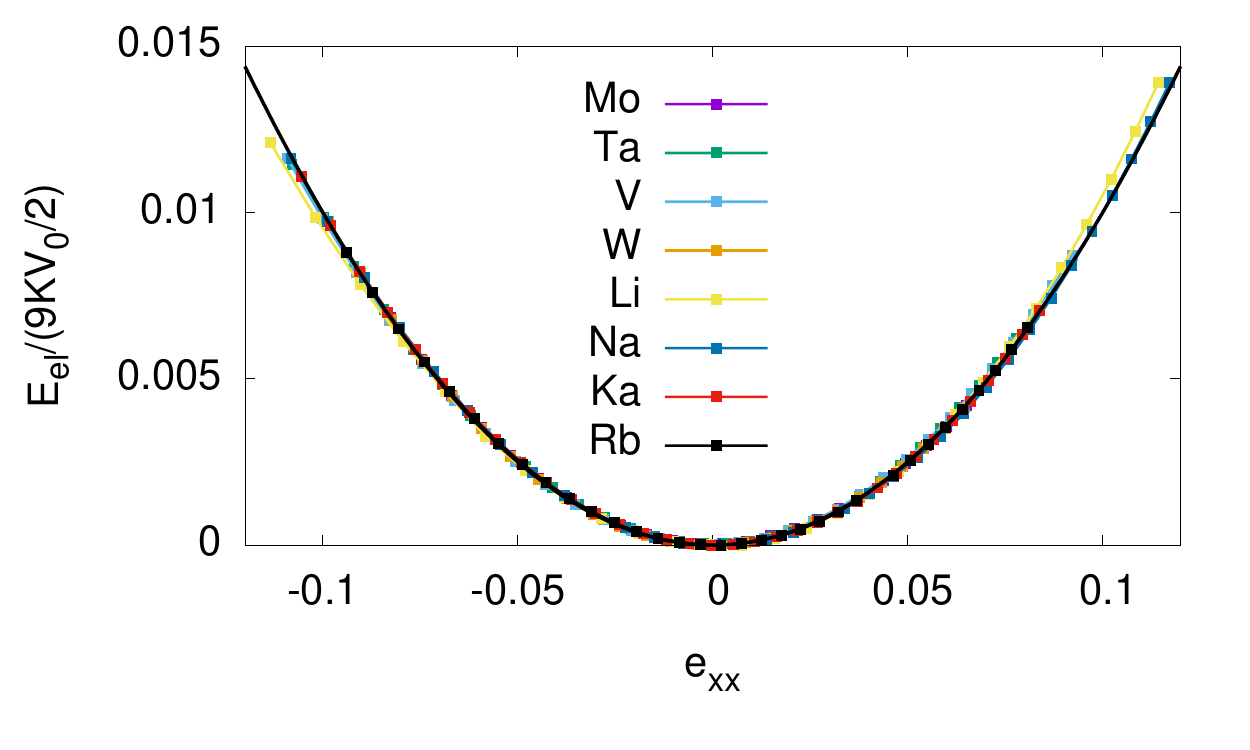}
\caption{(Color online) Elastic energy as a function of the isotropic Eulerian strain for various elemental metals, normalised to the bulk modulus. All data collapses to a simple Master curve $E=9 K V_0 e_{xx}^2/2$, which is equivalent to the Birch-Murnaghan model with $K'=4$ (continuous black curve). Lithium slightly deviates from this curve and has a value of $K'\approx 3.5$.}
\label{figabinitio2}
\end{center}
\end{figure}

To shed light on the specific value $K'=4$, which is suggested by the three-dimensional PFC model, we performed nonmagnetic {\em ab initio} simulations for various bcc elements.
$T=0 K$ results for the total energy as a function of the lattice constant are shown in Fig.~\ref{figabinitio1}.
If we present this data as a function of the Eulerian strain, we find that it becomes symmetric for many metals apart from lithium.
This symmetry corresponds to $K'=4$ in the Murnaghan models, in agreement with the PFC prediction.
The exception Li has a slight asymmetry and a value of $K'\approx 3.5$.
We note that the fitted values of $K'$ have an uncertainty, as can be seen from the (small) difference between Li and the other elements.
The main conclusion is that the shown bcc elements essentially lead to a parabolic curve if represented in terms of the Euler strain, which means that $K'$ is at least not too far from $K'=4$. 
Fitting the bulk modulus from the curvature near the minimum therefore allows to reduce all data (apart from Li) to one curve, see Fig.~\ref{figabinitio2}.
In essence, we can therefore conclude that the PFC and amplitude equations models, which predict $K'=4$ for bcc, are able to capture well the low temperature non-linear elasticity for various elements. 
For these low temperature applications, a large value of $\epsilon\gg\bar{\Delta}$ has to be chosen, such that the amplitudes are essentially strain independent.
By adjusting the value of $q_0$ according to the equilibrium lattice constant, $q_0=2\pi/a_0$, and multiplying the phase field crystal energy with a dimensional energy prefactor, one matches the bulk modulus of each element.

\begin{figure}
\begin{center}
\includegraphics[width=8.5cm]{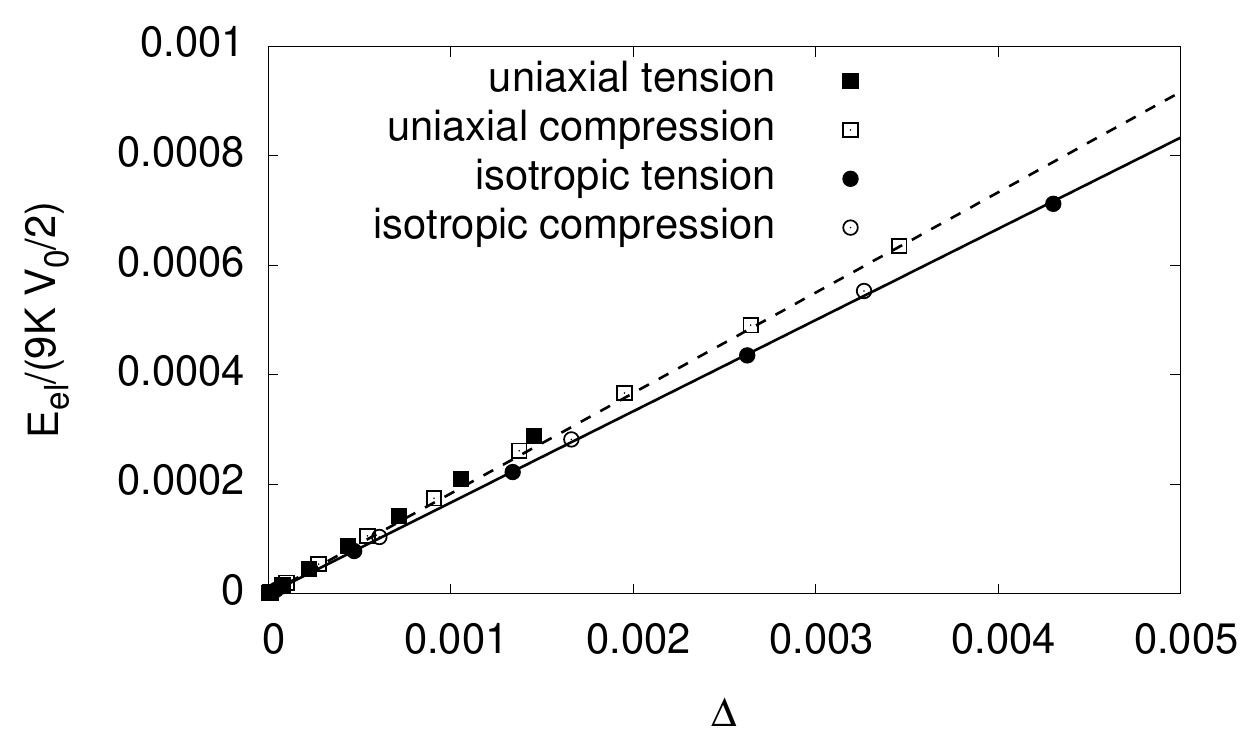}
\caption{Uniaxial and isotropic straining of tungsten.
The normalized elastic energy is shown versus the parameter $\Delta=\bar{\Delta}$ for these loadings.
The {\em ab initio} data falls on a straight line for compression and tension for isotropic loading;
the theoretical prediction is the solid line.
For uniaxial loading the data also collapses onto a straight (dashed) line both for tension and compression, but exhibits a slightly different slope due to violations of the Cauchy relation.
}
\label{fig11}
\end{center}
\end{figure}

We can exploit the comparison between the phase field crystal and the {\em ab initio} calculations even further.
For uniaxial stretching in [100] direction we can again predict the elastic energy and compare it to the $T=0$ {\em ab initio} results.
The continuum theory predicts that the energy should be linear in $\Delta=e_{xx}^2=\bar{\Delta}=\bar{e}_{xx}^2$ in the low temperature regime.
The {\em ab initio} data for tungsten fully confirms this expectation, see Fig.~\ref{fig11}.
Also, the behavior is symmetric under tension and compression in this representation, as the data (open and filled squares) falls onto a single straight line.
Moreover, from the definition of $\bar{\Delta}$ in Eq.~(\ref{eq49}), we expect that the elastic energy for the isotropic stretching should be six times larger than for the uniaxial stretching for the same value of $e_{xx}$, as then $\bar{\Delta}=6\bar{e}_{xx}^2$.
Fig.~\ref{fig11} therefore also contains the previous data for tungsten for isotropic straining, both in the compressive and tensile regime (open and filled circles, respectively).
The describing straight line has a similar, but slightly different slope compared to the uniaxial case.
We attribute this to slight deviations from the Cauchy relation $C_{12}=C_{44}$ for tungsten;
this relation is exactly fulfilled in the PFC model\cite{Spatschek:2010fk}.
Still, the PFC model gives an excellent description also for this type of mechanical loading.


Analogous to the three-dimensional expression (\ref{Mur1}) we can propose a similar expression for two dimensions.
For that we start with the ansatz
\begin{equation} \label{Mur2Dguessed}
E_{2D} = 2 K_{2D} V_0 e_{xx}^2 [1+\alpha(\beta-K'_{2D})e_{xx}],
\end{equation}
where the undeformed two-dimensional volume is $V_0=a_0^2$, in comparison to the deformed volume $V=a^2$.
We note that the choice of the global prefactor $2$ is here a matter of choice and only rescales the two-dimensional bulk modulus $K_{2D}$, which is not in the focus of the present investigations.
The non-linear Euler strain is
\begin{equation}
e_{xx}=\bar{e}_{xx}=e_{yy}=\bar{e}_{yy}= \frac{1}{2} \left(1-\frac{V_0}{V} \right), 
\end{equation}
and all other strain components vanish for an isotropic deformation.
The coefficients $\alpha$ and $\beta$ in Eq.~(\ref{Mur2Dguessed}) are determined by the requirement that the zero strain bulk modulus derivative (\ref{thermointen3}) is recovered.
From this we get $\alpha=2/3$ and $\beta=5$.
This can be compared with the low temperature limit $\pfcepsilon\gg\Delta$ of the 2D PFC result (\ref{intermediateresult}), 
\begin{equation} \label{Mur2DPFC}
f_{\mathrm{PFC}, 2D} = \frac{1}{2} K_{2D} e_{xx}^2,
\end{equation}
where a term, which is cubic in the strain $e_{xx}$, does not appear for constant amplitudes.
Here we have identified $K_{2D}=24 |A_0|^2$.
These results therefore suggest $K'_{2D}=5$.

\begin{figure}
\begin{center}
\includegraphics[width=8cm]{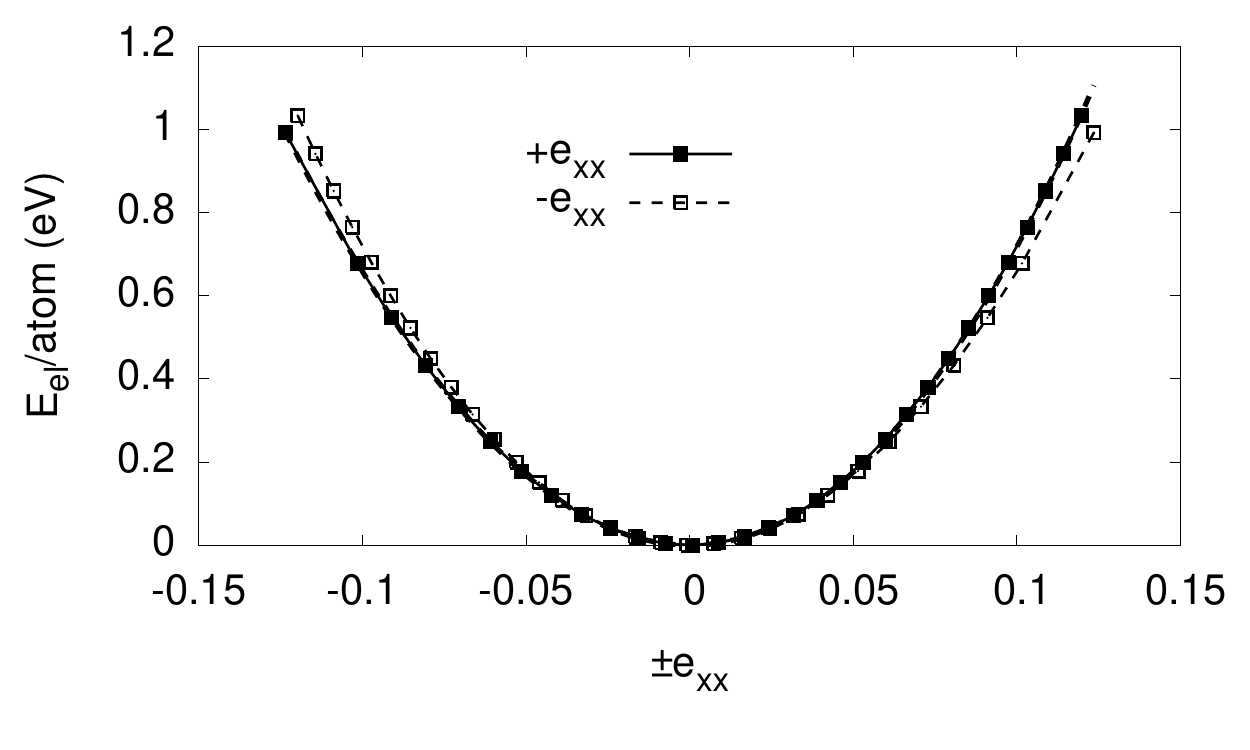}
\caption{Elastic energy of graphene, plotted versus the Euler strain, as obtained from the {\em ab initio} simulations. 
The functional form slightly deviates from a pure parabola and is well described by Eq.~(\ref{Mur2Dguessed}) with $K'_{2D}\approx 4.3$.
The filled symbols are the DFT data with the closed curve according to Eq.~(\ref{Mur2Dguessed}). 
The open symbols are the DFT data shown as a function of negative strain $e_{xx}$ to visualize the slight asymmetry and hence the deviation from a purely parabolic function.
}
\label{figgraphene}
\end{center}
\end{figure}

We have performed {\em ab initio} simulations of graphene for isotropic deformations, $e_{xx}=e_{yy}$. 
The data, plotted versus the Euler strain $e_{xx}$ is shown in Fig.~\ref{figgraphene}.
The functional form is again very close to the parabolic (and symmetric) form (\ref{Mur2DPFC}), but shows a small asymmetry, which can be fitted by $K'_{2D}\approx 4.3$ in Eq.~(\ref{Mur2Dguessed}).
The deviation may be due to the effect that the graphene structure deviates from the triangular structure of the 2D PFC model.
The extension of the analysis to the recent graphene model \cite{MSeymour:2016aa} may shed light on this issue.
Still, we can conclude from Fig.~\ref{figgraphene} that the (standard) PFC model gives a good description of the elastic response in a wide strain regime also for graphene.

\section{Summary and conclusions}
\label{SummaryConclusion}

We have analysed the non-linear elastic response of the phase field crystal in one, two and three dimensions for different crystal structures.
First, we have elaborated that the proper interpretation of (non-linear) elasticity is via the energy {\em density} in the PFC model, which has to be compared to the energy {\em per unit cell} for discrete atomistic descriptions.

A natural outcome of the differential operators in the amplitude equation formulation of the PFC model is the appearance of geometric non-linearity.
For elevated temperatures, additional physical non-linearity appears, which shows up via strain dependent amplitudes and can be understood as precursors of stress induced melting.

Both with and without physical non-linearity the response to deformation can be described through the non-linear strain tensor $\bar{e}_{ij}$ as given by Eq.~(\ref{strangeEulerStrain}), which is based on the right Cauchy-Green deformation tensor.
For the one- and two-dimensional case and isotropic loading the elastic response can equivalently be described through the Euler-Almansi tensor $e_{ij}$, but this is not the case for bcc ordering.
In general, the phase field crystal model has to be interpreted as a Eulerian description of elasticity.
A particular outcome is that the non-linear elastic energy depends on the strain tensor components $\bar{e}_{ij}$ in a symmetric way under compression and tension, as expressed through the dimensionless quantity $\bar{\Delta}$.

In the low temperature limit the PFC predictions for energy-volume curves coincide with the Birch-Murnaghan expression in three dimensions with bulk modulus derivative $K'=4$, and $K'_{2D}=5$ in two dimensions.
These suggested values are in good agreement with {\em ab initio} calculated energy-volume curves for various nonmagnetic bcc elements and graphene.
Also, other deformations like large uniaxial strains are well described by the PFC model.
We can therefore conclude that the phenomenological PFC model is well suitable to describe non-linear elastic deformation.

It is quite remarkable that the heuristic PFC model in its simplest form can capture a wide range of non-linear elastic response so well, as compared to electronic structure calculations.
It suggests that the effect of elastic deformations is already well described by the effective atom densities, which are the basis for the classical density functional theory and therefore the PFC model.
Moreover, the results indicate that the one-mode approximation is particularly good for the bcc elements.
It is known that the representation of e.g.~fcc requires to include more modes and reciprocal lattice vectors\cite{Wu:2010vn}.
The investigation of the non-linear elastic response for these cases will be subject of future research activities.

\begin{acknowledgments}
This work has been supported by the DFG priority program SPP 1713.
The authors gratefully acknowledge the computing time granted on the supercomputer JURECA at the J\"ulich Supercomputing Centre (JSC).
Furthermore, this research was supported by the Academy of Sciences
of the Czech Republic through the Fellowship of Jan Evangelista
Purkyn\v{e} (M.F.). The access to the computational resources
provided by the MetaCentrum under the program LM2010005
and the CERIT-SC under the program Center CERIT Scientific Cloud,
part of the Operational Program Research and Development for
Innovations, Reg. No. CZ.1.05/3.2.00/08.0144, is highly appreciated.
This work was also supported by the IT4Innovations Centre of Excellence
project (CZ.1.05/1.1.00/02.0070), funded by the European Regional
Development Fund and the national budget of the Czech Republic
via the Research and Development for Innovations Operational
Program, as well as Czech Ministry of Education, Youth and Sports
via the project Large Research, Development and Innovations
Infrastructures (LM2011033).

\end{acknowledgments}

\appendix

\section{Derivation of the amplitude equations}
\label{appendix::aederivation}

We use here a derivation of the amplitude equations which starts directly from the energy functional instead from the evolution equations, following the approach in Ref.~\onlinecite{Chan:2009aa}.
This offers a direct way to obtain the proper equations without the need to create a generating functional a posteriori \cite{Wu:2007kx}.
We also point out that the rotationally invariant different operator, which is frequently found in amplitude equations comes out automatically, without the need to justify it from more complex renormalization group theory approaches \cite{PhysRevLett.80.3888,PhysRevLett.76.2185}.
We briefly illustrate this for the two-dimensional triangular model and then give the result for the three-dimensional bcc case, which can be obtained similarly.

We start from the PFC energy functional (\ref{pfcfunctional})
for $q_0=1$.
The amplitude expansion is written according to Eq.~(\ref{onemode})
with the reciprocal lattice vectors given in (\ref{2DRLVs}).
To get the amplitude equation energy functional we insert the one-mode approximation into the PFC energy and assume that the amplitudes are varying on a scale much larger than $1/q_0$.
Then we get under the assumption that the system volume $V$ is a multiple of the lattice unit 
\begin{equation}
\int_V f(\{A_j\}) \exp(i\mathbf{k}\cdot\mathbf{r}) d\mathbf{r} = \int_V f(\{A_j\}) \delta_{\mathbf{k}, 0} d\mathbf{r}
\end{equation}
for any slow function $f(\{A_j\})$ of the amplitudes.
In particular, this step includes the averaging over unit cells.

For the quartic term
\begin{equation}
F_4 := \frac{1}{4} \int d\mathbf{r} \psi^4,
\end{equation}
only terms, which belong to a closed polygon of RLVs, contribute.
In the following we decompose $F_4=F_{44}+F_{43}+F_{42}+F_{40}$, where the second subscript denotes the order of the amplitudes.

For the contribution, which is quartic in the amplitudes $A_j$, there are two types of RLV configurations, namely $\mathbf{k}^{(i)}-\mathbf{k}^{(i)}+\mathbf{k}^{(i)}-\mathbf{k}^{(i)}=0$ and $\mathbf{k}^{(i)}-\mathbf{k}^{(i)}+\mathbf{k}^{(j)}-\mathbf{k}^{(j)}=0$ with $i\neq j$.
Forming all possible combinations we get for this
\begin{eqnarray}
F_{44}/V &=& \frac{1}{4} \Big( 6 |A_1|^4 + 6 |A_2|^4 + 6 |A_3|^4 +  \\
&=& 24|A_1|^2|A_2|^2 + 24|A_1|^2|A_3|^2 + 24|A_2|^2|A_3|^2 \Big). \nonumber
\end{eqnarray}
For each product of amplitudes, which leads to a closed polygon of reciprocal lattice vectors, we have to count the number of combinations with which it appears in the term proportional to $\psi^4$.
For the products, which consist of $A_j^2 {A_j^*}^2$ we therefore have to arrange all possible combinations of these four factors.
The coefficient 6 appears, because there are $4\cdot 3/2$ possibilities for placing the amplitude $A_j$ twice, and the remaining two places in a product are occupied by $A_j^*$.
For the mixed terms we get $4\cdot 3 \cdot 2 \cdot 1=24$ possibilities to arrange $A_1$, $A_1^*$, $A_2$ and $A_2^*$.
This completes the calculation of the quartic terms.

The cubic terms are also generated by $F_4$.
Here the only closed polygon is $\mathbf{k}^{(1)}+\mathbf{k}^{(2)}+\mathbf{k}^{(3)}=0$, hence we get terms containing $A_1A_2A_3$ and $A_1^*A_2^*A_3^*$.
Notice that one of the amplitudes ($A_1$) is negative, hence these products are negative for positive $\bar{\psi}$, see Eq.~(\ref{2Damplitudesigns}).
This is necessary to stabilise the solid phase.
We get
\begin{equation}
F_{43}/V = \frac{1}{4} \bar{\psi} \left( 24 A_1A_2A_3 + 24 A_1^*A_2^*A_3^* \right).
\end{equation}
We have $4 \cdot 3 \cdot 2 \cdot 1$ possibilities to arrange the four factors $A_1$, $A_2$, $A_3$, $\bar{\psi}$ or $A_1^*$, $A_2^*$, $A_3^*$, $\bar{\psi}$.

The quadratic terms stem from the $\psi^4$, $\psi^2$ and the gradient term.
We start with $F_{42}$.
Only combinations of antiparallel reciprocal lattice vectors contribute here.
There we get
\begin{equation}
F_{42}/V = \frac{1}{4}\bar{\psi}^2 \left( 12|A_1|^2 + 12|A_2|^2 + 12|A_3|^2\right),
\end{equation}
with $12=4\cdot3\cdot2/2$ for choosing the positions of the individual factors in a product of $2\times\bar{\psi}$, $A$, $A^*$.

There is no contribution of the type $F_{41}$, as the oscillating factors cancel.
However, there is a term $F_{40}$, which only involves the constant contributions,
\begin{equation}
F_{40}/V = \frac{1}{4} \bar{\psi}^4.
\end{equation}

The local quadratic energy is defined as
\begin{equation}
F_2 := \frac{1}{2} (1-\pfcepsilon)\int d\mathbf{r} \psi^2.
\end{equation}
It gives
\begin{equation}
F_2/V = \frac{1}{2} (1-\pfcepsilon) \left( 2 |A_1|^2+2 |A_2|^2+ 2|A_3|^2 + \bar{\psi}^2\right).
\end{equation}

The most difficult term is the one which contains gradients of $\psi$.
It leads both to terms which are local and nonlocal in the amplitudes.
It is defined as
\begin{eqnarray}
F_{grad} &=& \int f_{grad} d\mathbf{r} = \int \left( \frac{\psi}{2} (\nabla^4\psi + 2\nabla^2\psi) \right) d\mathbf{r} \nonumber \\
&=& F_{grad, A} + F_{grad, \nabla A}.
\end{eqnarray}
The term $F_{grad, A}$ contains only the terms which are local in $A$, because the differentiation acts on the exponential term in the product rule.
Then, each differentiation simply brings down a factor $\pm i\mathbf{k}$, and we get with $|\mathbf{k}|=1$
\begin{eqnarray}
F_{grad, A} &=& \int \bigg[ 2\times\frac{1}{2} \left( |A_1|^2 + |A_2|^2 + |A_3|^2 \right) \nonumber \\
&& - 2\times \left( |A_1|^2 + |A_2|^2 + |A_3|^2 \right)  \bigg] d\mathbf{r} \nonumber \\
&=& -\int  \left( |A_1|^2 + |A_2|^2 + |A_3|^2 \right) d\mathbf{r},
\end{eqnarray}
where again we retained only the term which do not contain fast oscillating factors.
The factor $2\times$ comes from the fact that each combination $A_iA_i^*$ can be obtained with either $A_i^*$ or $A_i$ being in front in a product $\psi^2$.

Altogether, the local terms therefore form the ``double well potential'', in analogy to classical phase field models.
It is given by
\begin{eqnarray}
F_{dw} &=& F_{44} + F_{43} + F_{42} + F_{40} + F_2 + F_{grad, A} , \nonumber \\
&=& \int d\mathbf{r} \Bigg\{ \frac{3}{2} |A_1|^4 + \frac{3}{2} |A_2|^4 + \frac{3}{2} |A_3|^4 + 6|A_1|^2|A_2|^2 \nonumber \\
&& + 6|A_1|^2|A_3|^2 + 6|A_2|^2|A_3|^2 \nonumber \\
&&+ \bar{\psi} \left( 6 A_1A_2A_3 + 6 A_1^*A_2^*A_3^* \right) \nonumber \\
&&+ (3\bar{\psi}^2-\pfcepsilon)  \left( |A_1|^2 + |A_2|^2 + |A_3|^2 \right) \nonumber \\
&& + \frac{1}{2}(1-\pfcepsilon)\bar{\psi}^2 + \frac{1}{4} \bar{\psi}^4 \Bigg\},
\end{eqnarray}
which coincides with Chan's and Goldenfeld's result\cite{Chan:2009aa}, apart from the terms independent of the amplitudes.
We point out that they are not relevant for the amplitude dynamics, as they vanish during the variational derivative.
However, these terms still influence the energy and are therefore required to compare the PFC energy (\ref{pfcfunctional}) with the one expressed through the amplitudes.

For evaluating the nonlocal terms in $A$ we perform an integration by part of $F_{grad}$ and retain afterwards only the derivatives acting on $A$, not on the exponential factor, which are already covered by $F_{grad, A}$ (the previous local contribution is the same whether we use the integration by part or not).
Hence
\begin{equation}
F_{grad} = \int \left( \frac{1}{2} (\nabla^2\psi)^2 - (\nabla\psi)^2  \right) d\mathbf{r}.
\end{equation}
From this we get
\begin{equation}
F_{grad, \nabla A} = \int \sum_{j=1}^3 \left| (\nabla^2 + 2i \mathbf{k}^{(j)}\cdot\nabla)A_j \right|^2 d\mathbf{r},
\end{equation}
where again we retain only the terms which contain gradients of the amplitudes.
With ${L}_j=\nabla^2+2i\mathbf{k}^{(j)}\cdot\nabla$ therefore altogether
\begin{eqnarray}
F &=& \int d\mathbf{r} \Bigg\{ \sum_{j=1}^3 \left| {L}_j A_j \right|^2 + \frac{3}{2} |A_1|^4 + \frac{3}{2} |A_2|^4 + \frac{3}{2} |A_3|^4 \nonumber \\
&&+ 6|A_1|^2|A_2|^2 + 6|A_1|^2|A_3|^2 + 6|A_2|^2|A_3|^2 \nonumber \\
&&+ \bar{\psi} \left( 6 A_1A_2A_3 + 6 A_1^*A_2^*A_3^* \right) \nonumber \\
&&+ (3\bar{\psi}^2-\pfcepsilon)  \left( |A_1|^2 + |A_2|^2 + |A_3|^2 \right) \nonumber \\
&& + \frac{1}{2}(1-\pfcepsilon)\bar{\psi}^2 + \frac{1}{4} \bar{\psi}^4 \Bigg\}.
\end{eqnarray}
Apart from the last two amplitude independent terms this expression is the same as in Eq.~(\ref{eq22}).
Alternatively, we express the differential operator as ${L}_j = 2i\,\Box_j$ using
\begin{equation}
\Box_j=\mathbf{k}^{(j)}\cdot\nabla - \frac{i}{2q_0}\nabla^2
\end{equation}
with $q_0=|\mathbf{k}^{(j)}|$.

For the three-dimensional bcc model we can proceed in the same way.
We start from the same phase field crystal model, but this time with the reciprocal lattice vectors given in (\ref{3DRLVs}).
Notice that $\bar{\psi}$ is assumed to be negative there, in agreement with the calculation by Wu and Karma\cite{Wu:2006uq,Wu:2007kx}.
By inserting the amplitude expansion into the functional and integrating over multiples of the unit cells we obtain similarly to above the functional (\ref{nonlinAE}).

\bibliography{references}

\begin{thebibliography}{36}%
\makeatletter
\providecommand \@ifxundefined [1]{%
 \@ifx{#1\undefined}
}%
\providecommand \@ifnum [1]{%
 \ifnum #1\expandafter \@firstoftwo
 \else \expandafter \@secondoftwo
 \fi
}%
\providecommand \@ifx [1]{%
 \ifx #1\expandafter \@firstoftwo
 \else \expandafter \@secondoftwo
 \fi
}%
\providecommand \natexlab [1]{#1}%
\providecommand \enquote  [1]{``#1''}%
\providecommand \bibnamefont  [1]{#1}%
\providecommand \bibfnamefont [1]{#1}%
\providecommand \citenamefont [1]{#1}%
\providecommand \href@noop [0]{\@secondoftwo}%
\providecommand \href [0]{\begingroup \@sanitize@url \@href}%
\providecommand \@href[1]{\@@startlink{#1}\@@href}%
\providecommand \@@href[1]{\endgroup#1\@@endlink}%
\providecommand \@sanitize@url [0]{\catcode `\\12\catcode `\$12\catcode
  `\&12\catcode `\#12\catcode `\^12\catcode `\_12\catcode `\%12\relax}%
\providecommand \@@startlink[1]{}%
\providecommand \@@endlink[0]{}%
\providecommand \url  [0]{\begingroup\@sanitize@url \@url }%
\providecommand \@url [1]{\endgroup\@href {#1}{\urlprefix }}%
\providecommand \urlprefix  [0]{URL }%
\providecommand \Eprint [0]{\href }%
\providecommand \doibase [0]{http://dx.doi.org/}%
\providecommand \selectlanguage [0]{\@gobble}%
\providecommand \bibinfo  [0]{\@secondoftwo}%
\providecommand \bibfield  [0]{\@secondoftwo}%
\providecommand \translation [1]{[#1]}%
\providecommand \BibitemOpen [0]{}%
\providecommand \bibitemStop [0]{}%
\providecommand \bibitemNoStop [0]{.\EOS\space}%
\providecommand \EOS [0]{\spacefactor3000\relax}%
\providecommand \BibitemShut  [1]{\csname bibitem#1\endcsname}%
\let\auto@bib@innerbib\@empty
\bibitem [{\citenamefont {Elder}\ and\ \citenamefont
  {Grant}(2004)}]{Elder:2004ys}%
  \BibitemOpen
  \bibfield  {author} {\bibinfo {author} {\bibfnamefont {K.~R.}\ \bibnamefont
  {Elder}}\ and\ \bibinfo {author} {\bibfnamefont {M.}~\bibnamefont {Grant}},\
  }\href@noop {} {\bibfield  {journal} {\bibinfo  {journal} {Phys. Rev. E}\
  }\textbf {\bibinfo {volume} {70}},\ \bibinfo {pages} {051605} (\bibinfo
  {year} {2004})}\BibitemShut {NoStop}%
\bibitem [{\citenamefont {Elder}\ \emph {et~al.}(2002)\citenamefont {Elder},
  \citenamefont {Katakowski}, \citenamefont {Haataja},\ and\ \citenamefont
  {Grant}}]{PhysRevLett.88.245701}%
  \BibitemOpen
  \bibfield  {author} {\bibinfo {author} {\bibfnamefont {K.~R.}\ \bibnamefont
  {Elder}}, \bibinfo {author} {\bibfnamefont {M.}~\bibnamefont {Katakowski}},
  \bibinfo {author} {\bibfnamefont {M.}~\bibnamefont {Haataja}}, \ and\
  \bibinfo {author} {\bibfnamefont {M.}~\bibnamefont {Grant}},\ }\href
  {\doibase 10.1103/PhysRevLett.88.245701} {\bibfield  {journal} {\bibinfo
  {journal} {Phys. Rev. Lett.}\ }\textbf {\bibinfo {volume} {88}},\ \bibinfo
  {pages} {245701} (\bibinfo {year} {2002})}\BibitemShut {NoStop}%
\bibitem [{\citenamefont {Heinonen}\ \emph {et~al.}(2016)\citenamefont
  {Heinonen}, \citenamefont {Achim}, \citenamefont {Kosterlitz}, \citenamefont
  {Ying}, \citenamefont {Lowengrub},\ and\ \citenamefont
  {Ala-Nissila}}]{HeinonenV:2016prl}%
  \BibitemOpen
  \bibfield  {author} {\bibinfo {author} {\bibfnamefont {V.}~\bibnamefont
  {Heinonen}}, \bibinfo {author} {\bibfnamefont {C.~V.}\ \bibnamefont {Achim}},
  \bibinfo {author} {\bibfnamefont {J.~M.}\ \bibnamefont {Kosterlitz}},
  \bibinfo {author} {\bibfnamefont {S.-C.}\ \bibnamefont {Ying}}, \bibinfo
  {author} {\bibfnamefont {J.}~\bibnamefont {Lowengrub}}, \ and\ \bibinfo
  {author} {\bibfnamefont {T.}~\bibnamefont {Ala-Nissila}},\ }\href@noop {}
  {\bibfield  {journal} {\bibinfo  {journal} {Phys. Rev. Lett.}\ }\textbf
  {\bibinfo {volume} {116}},\ \bibinfo {pages} {024303} (\bibinfo {year}
  {2016})}\BibitemShut {NoStop}%
\bibitem [{\citenamefont {Kocher}\ and\ \citenamefont
  {Provatas}(2015)}]{KocherG:2015prl}%
  \BibitemOpen
  \bibfield  {author} {\bibinfo {author} {\bibfnamefont {G.}~\bibnamefont
  {Kocher}}\ and\ \bibinfo {author} {\bibfnamefont {N.}~\bibnamefont
  {Provatas}},\ }\href@noop {} {\bibfield  {journal} {\bibinfo  {journal}
  {Phys. Rev. Lett.}\ }\textbf {\bibinfo {volume} {114}},\ \bibinfo {pages}
  {155501} (\bibinfo {year} {2015})}\BibitemShut {NoStop}%
\bibitem [{\citenamefont {Tarp}\ \emph {et~al.}(2014)\citenamefont {Tarp},
  \citenamefont {Angheluta}, \citenamefont {Mathiesen},\ and\ \citenamefont
  {Goldenfeld}}]{TarpJM:2014prl}%
  \BibitemOpen
  \bibfield  {author} {\bibinfo {author} {\bibfnamefont {J.~M.}\ \bibnamefont
  {Tarp}}, \bibinfo {author} {\bibfnamefont {L.}~\bibnamefont {Angheluta}},
  \bibinfo {author} {\bibfnamefont {J.}~\bibnamefont {Mathiesen}}, \ and\
  \bibinfo {author} {\bibfnamefont {N.}~\bibnamefont {Goldenfeld}},\
  }\href@noop {} {\bibfield  {journal} {\bibinfo  {journal} {Phys. Rev. Lett.}\
  }\textbf {\bibinfo {volume} {113}},\ \bibinfo {pages} {265503} (\bibinfo
  {year} {2014})}\BibitemShut {NoStop}%
\bibitem [{\citenamefont {Berry}\ and\ \citenamefont
  {Grant}(2011)}]{BerryJ:2011prl}%
  \BibitemOpen
  \bibfield  {author} {\bibinfo {author} {\bibfnamefont {J.}~\bibnamefont
  {Berry}}\ and\ \bibinfo {author} {\bibfnamefont {M.}~\bibnamefont {Grant}},\
  }\href@noop {} {\bibfield  {journal} {\bibinfo  {journal} {Phys. Rev. Lett}\
  }\textbf {\bibinfo {volume} {106}},\ \bibinfo {pages} {175702} (\bibinfo
  {year} {2011})}\BibitemShut {NoStop}%
\bibitem [{\citenamefont {Seymour}\ and\ \citenamefont
  {Provatas}(2016)}]{MSeymour:2016aa}%
  \BibitemOpen
  \bibfield  {author} {\bibinfo {author} {\bibfnamefont {M.}~\bibnamefont
  {Seymour}}\ and\ \bibinfo {author} {\bibfnamefont {N.}~\bibnamefont
  {Provatas}},\ }\href@noop {} {\bibfield  {journal} {\bibinfo  {journal}
  {Phys. Rev. B}\ }\textbf {\bibinfo {volume} {93}},\ \bibinfo {pages} {035447}
  (\bibinfo {year} {2016})}\BibitemShut {NoStop}%
\bibitem [{\citenamefont {Greenwood}\ \emph {et~al.}(2010)\citenamefont
  {Greenwood}, \citenamefont {Provatas},\ and\ \citenamefont
  {Rottler}}]{GreenwoodM:2010prl}%
  \BibitemOpen
  \bibfield  {author} {\bibinfo {author} {\bibfnamefont {M.}~\bibnamefont
  {Greenwood}}, \bibinfo {author} {\bibfnamefont {N.}~\bibnamefont {Provatas}},
  \ and\ \bibinfo {author} {\bibfnamefont {J.}~\bibnamefont {Rottler}},\
  }\href@noop {} {\bibfield  {journal} {\bibinfo  {journal} {Phys. Rev. Lett.}\
  }\textbf {\bibinfo {volume} {105}},\ \bibinfo {pages} {045702} (\bibinfo
  {year} {2010})}\BibitemShut {NoStop}%
\bibitem [{\citenamefont {Chan}\ and\ \citenamefont
  {Goldenfeld}(2009)}]{Chan:2009aa}%
  \BibitemOpen
  \bibfield  {author} {\bibinfo {author} {\bibfnamefont {P.~Y.}\ \bibnamefont
  {Chan}}\ and\ \bibinfo {author} {\bibfnamefont {N.}~\bibnamefont
  {Goldenfeld}},\ }\href@noop {} {\bibfield  {journal} {\bibinfo  {journal}
  {Phys. Rev. E}\ }\textbf {\bibinfo {volume} {80}},\ \bibinfo {pages}
  {065105(R)} (\bibinfo {year} {2009})}\BibitemShut {NoStop}%
\bibitem [{\citenamefont {H\"uter}\ \emph {et~al.}(2015)\citenamefont
  {H\"uter}, \citenamefont {Neugebauer}, \citenamefont {Boussinot},
  \citenamefont {Svendsen}, \citenamefont {Prahl},\ and\ \citenamefont
  {Spatschek}}]{Huter:2015aa}%
  \BibitemOpen
  \bibfield  {author} {\bibinfo {author} {\bibfnamefont {C.}~\bibnamefont
  {H\"uter}}, \bibinfo {author} {\bibfnamefont {J.}~\bibnamefont {Neugebauer}},
  \bibinfo {author} {\bibfnamefont {G.}~\bibnamefont {Boussinot}}, \bibinfo
  {author} {\bibfnamefont {B.}~\bibnamefont {Svendsen}}, \bibinfo {author}
  {\bibfnamefont {U.}~\bibnamefont {Prahl}}, \ and\ \bibinfo {author}
  {\bibfnamefont {R.}~\bibnamefont {Spatschek}},\ }\href@noop {} {\bibfield
  {journal} {\bibinfo  {journal} {Continuum Mech. Thermodyn.}\ ,\ \bibinfo
  {pages} {DOI 10.1007/s00161}} (\bibinfo {year} {2015})}\BibitemShut {NoStop}%
\bibitem [{\citenamefont {Singh}(1991)}]{Singh1991351}%
  \BibitemOpen
  \bibfield  {author} {\bibinfo {author} {\bibfnamefont {Y.}~\bibnamefont
  {Singh}},\ }\href {\doibase http://dx.doi.org/10.1016/0370-1573(91)90097-6}
  {\bibfield  {journal} {\bibinfo  {journal} {Physics Reports}\ }\textbf
  {\bibinfo {volume} {207}},\ \bibinfo {pages} {351 } (\bibinfo {year}
  {1991})}\BibitemShut {NoStop}%
\bibitem [{\citenamefont {Laird}\ \emph {et~al.}(1987)\citenamefont {Laird},
  \citenamefont {McCoy},\ and\ \citenamefont {Haymet}}]{ISI:A1987K583500046}%
  \BibitemOpen
  \bibfield  {author} {\bibinfo {author} {\bibfnamefont {B.~B.}\ \bibnamefont
  {Laird}}, \bibinfo {author} {\bibfnamefont {J.~D.}\ \bibnamefont {McCoy}}, \
  and\ \bibinfo {author} {\bibfnamefont {A.~D.~J.}\ \bibnamefont {Haymet}},\
  }\href {\doibase {10.1063/1.453663}} {\bibfield  {journal} {\bibinfo
  {journal} {J. Chem. Phys.}\ }\textbf {\bibinfo {volume} {87}},\ \bibinfo
  {pages} {5449} (\bibinfo {year} {1987})}\BibitemShut {NoStop}%
\bibitem [{\citenamefont {Spatschek}\ and\ \citenamefont
  {Karma}(2010)}]{Spatschek:2010fk}%
  \BibitemOpen
  \bibfield  {author} {\bibinfo {author} {\bibfnamefont {R.}~\bibnamefont
  {Spatschek}}\ and\ \bibinfo {author} {\bibfnamefont {A.}~\bibnamefont
  {Karma}},\ }\href@noop {} {\bibfield  {journal} {\bibinfo  {journal} {Phys.
  Rev. B}\ }\textbf {\bibinfo {volume} {81}},\ \bibinfo {pages} {214201}
  (\bibinfo {year} {2010})}\BibitemShut {NoStop}%
\bibitem [{\citenamefont {Wu}\ \emph {et~al.}(2006)\citenamefont {Wu},
  \citenamefont {Karma}, \citenamefont {Hoyt},\ and\ \citenamefont
  {Asta}}]{Wu:2006uq}%
  \BibitemOpen
  \bibfield  {author} {\bibinfo {author} {\bibfnamefont {K.-A.}\ \bibnamefont
  {Wu}}, \bibinfo {author} {\bibfnamefont {A.}~\bibnamefont {Karma}}, \bibinfo
  {author} {\bibfnamefont {J.~J.}\ \bibnamefont {Hoyt}}, \ and\ \bibinfo
  {author} {\bibfnamefont {M.}~\bibnamefont {Asta}},\ }\href@noop {} {\bibfield
   {journal} {\bibinfo  {journal} {Phys. Rev. B}\ }\textbf {\bibinfo {volume}
  {73}},\ \bibinfo {pages} {094101} (\bibinfo {year} {2006})}\BibitemShut
  {NoStop}%
\bibitem [{\citenamefont {Wu}\ and\ \citenamefont {Karma}(2007)}]{Wu:2007kx}%
  \BibitemOpen
  \bibfield  {author} {\bibinfo {author} {\bibfnamefont {K.-A.}\ \bibnamefont
  {Wu}}\ and\ \bibinfo {author} {\bibfnamefont {A.}~\bibnamefont {Karma}},\
  }\href@noop {} {\bibfield  {journal} {\bibinfo  {journal} {Phys. Rev. B}\
  }\textbf {\bibinfo {volume} {76}},\ \bibinfo {pages} {184107} (\bibinfo
  {year} {2007})}\BibitemShut {NoStop}%
\bibitem [{\citenamefont {Adland}\ \emph {et~al.}(2013)\citenamefont {Adland},
  \citenamefont {Karma}, \citenamefont {Spatschek}, \citenamefont {Buta},\ and\
  \citenamefont {Asta}}]{Adland:2013ys}%
  \BibitemOpen
  \bibfield  {author} {\bibinfo {author} {\bibfnamefont {A.}~\bibnamefont
  {Adland}}, \bibinfo {author} {\bibfnamefont {A.}~\bibnamefont {Karma}},
  \bibinfo {author} {\bibfnamefont {R.}~\bibnamefont {Spatschek}}, \bibinfo
  {author} {\bibfnamefont {D.}~\bibnamefont {Buta}}, \ and\ \bibinfo {author}
  {\bibfnamefont {M.}~\bibnamefont {Asta}},\ }\href@noop {} {\bibfield
  {journal} {\bibinfo  {journal} {Phys. Rev. B}\ }\textbf {\bibinfo {volume}
  {87}},\ \bibinfo {pages} {024110} (\bibinfo {year} {2013})}\BibitemShut
  {NoStop}%
\bibitem [{\citenamefont {Spatschek}\ \emph {et~al.}(2013)\citenamefont
  {Spatschek}, \citenamefont {Adland},\ and\ \citenamefont {Karma}}]{kar13}%
  \BibitemOpen
  \bibfield  {author} {\bibinfo {author} {\bibfnamefont {R.}~\bibnamefont
  {Spatschek}}, \bibinfo {author} {\bibfnamefont {A.}~\bibnamefont {Adland}}, \
  and\ \bibinfo {author} {\bibfnamefont {A.}~\bibnamefont {Karma}},\
  }\href@noop {} {\bibfield  {journal} {\bibinfo  {journal} {Physical Review
  B}\ }\textbf {\bibinfo {volume} {87}},\ \bibinfo {pages} {024109} (\bibinfo
  {year} {2013})}\BibitemShut {NoStop}%
\bibitem [{\citenamefont {Lejaeghere}\ \emph {et~al.}(2013)\citenamefont
  {Lejaeghere}, \citenamefont {Speybrock}, \citenamefont {Oost},\ and\
  \citenamefont {Cottenier}}]{LejaeghereK:2013crit}%
  \BibitemOpen
  \bibfield  {author} {\bibinfo {author} {\bibfnamefont {K.}~\bibnamefont
  {Lejaeghere}}, \bibinfo {author} {\bibfnamefont {V.~V.}\ \bibnamefont
  {Speybrock}}, \bibinfo {author} {\bibfnamefont {G.~V.}\ \bibnamefont {Oost}},
  \ and\ \bibinfo {author} {\bibfnamefont {S.}~\bibnamefont {Cottenier}},\
  }\href@noop {} {\bibfield  {journal} {\bibinfo  {journal} {Critical Reviews
  in Solid State and Materials Science}\ }\textbf {\bibinfo {volume} {39}},\
  \bibinfo {pages} {1} (\bibinfo {year} {2013})}\BibitemShut {NoStop}%
\bibitem [{\citenamefont {Landau}\ and\ \citenamefont
  {Lifshitz}(1987)}]{LandauLifshitz:7}%
  \BibitemOpen
  \bibfield  {author} {\bibinfo {author} {\bibfnamefont {L.~D.}\ \bibnamefont
  {Landau}}\ and\ \bibinfo {author} {\bibfnamefont {E.~M.}\ \bibnamefont
  {Lifshitz}},\ }\href@noop {} {\emph {\bibinfo {title} {Elasticity theory}}},\
  edited by\ \bibinfo {editor} {\bibfnamefont {H.~G.}\ \bibnamefont
  {Sch{\"o}pf}},\ Vol.~\bibinfo {volume} {7}\ (\bibinfo  {publisher} {Akademie
  Verlag},\ \bibinfo {year} {1987})\BibitemShut {NoStop}%
\bibitem [{\citenamefont {Nemat-Nasser}\ and\ \citenamefont
  {Hori}(1999)}]{NematNasserHori:1}%
  \BibitemOpen
  \bibfield  {author} {\bibinfo {author} {\bibfnamefont {S.}~\bibnamefont
  {Nemat-Nasser}}\ and\ \bibinfo {author} {\bibfnamefont {M.}~\bibnamefont
  {Hori}},\ }\href@noop {} {\emph {\bibinfo {title} {{M}icromechanics: overall
  properties of heterogenous materials}}}\ (\bibinfo  {publisher} {Elsevier
  North-Holland},\ \bibinfo {year} {1999})\BibitemShut {NoStop}%
\bibitem [{\citenamefont {Dimitrienko}(2011)}]{DimitrienkoY:2011b}%
  \BibitemOpen
  \bibfield  {author} {\bibinfo {author} {\bibfnamefont {Y.}~\bibnamefont
  {Dimitrienko}},\ }\href@noop {} {\emph {\bibinfo {title} {Nonlinear Continuum
  Mechanics and Large Inelastic Deformations}}},\ \bibinfo {series} {Series:
  Solid Mechanics and Its Applications}, Vol.\ \bibinfo {volume} {174}\
  (\bibinfo  {publisher} {Springer Netherlands},\ \bibinfo {year}
  {2011})\BibitemShut {NoStop}%
\bibitem [{\citenamefont {Clayton}(2014)}]{Clayton:2014aa}%
  \BibitemOpen
  \bibfield  {author} {\bibinfo {author} {\bibfnamefont {J.}~\bibnamefont
  {Clayton}},\ }\href@noop {} {\bibfield  {journal} {\bibinfo  {journal} {Int.
  J. Appl. Mech.}\ }\textbf {\bibinfo {volume} {6}},\ \bibinfo {pages}
  {1450048} (\bibinfo {year} {2014})}\BibitemShut {NoStop}%
\bibitem [{\citenamefont {Chan}\ \emph {et~al.}(2009)\citenamefont {Chan},
  \citenamefont {Goldenfeld},\ and\ \citenamefont {Dantzig}}]{ChanPY:2009pre}%
  \BibitemOpen
  \bibfield  {author} {\bibinfo {author} {\bibfnamefont {P.~Y.}\ \bibnamefont
  {Chan}}, \bibinfo {author} {\bibfnamefont {N.}~\bibnamefont {Goldenfeld}}, \
  and\ \bibinfo {author} {\bibfnamefont {J.}~\bibnamefont {Dantzig}},\
  }\href@noop {} {\bibfield  {journal} {\bibinfo  {journal} {Phys. Rev. E}\
  }\textbf {\bibinfo {volume} {79}},\ \bibinfo {pages} {035701} (\bibinfo
  {year} {2009})}\BibitemShut {NoStop}%
\bibitem [{\citenamefont {Eckhaus}(1965)}]{EckhausW:1965b}%
  \BibitemOpen
  \bibfield  {author} {\bibinfo {author} {\bibfnamefont {W.}~\bibnamefont
  {Eckhaus}},\ }\href@noop {} {\emph {\bibinfo {title} {Studies in Non-Linear
  Stability Theory}}},\ \bibinfo {series} {Springer Tracts in natural
  philosophy}, Vol.~\bibinfo {volume} {6}\ (\bibinfo  {publisher} {Springer
  Verlag New York},\ \bibinfo {year} {1965})\BibitemShut {NoStop}%
\bibitem [{\citenamefont {Jaatinen}\ \emph {et~al.}(2009)\citenamefont
  {Jaatinen}, \citenamefont {Achim}, \citenamefont {Elder},\ and\ \citenamefont
  {Ala-Nissila}}]{JaatinenA:2009pre}%
  \BibitemOpen
  \bibfield  {author} {\bibinfo {author} {\bibfnamefont {A.}~\bibnamefont
  {Jaatinen}}, \bibinfo {author} {\bibfnamefont {C.~V.}\ \bibnamefont {Achim}},
  \bibinfo {author} {\bibfnamefont {K.~R.}\ \bibnamefont {Elder}}, \ and\
  \bibinfo {author} {\bibfnamefont {T.}~\bibnamefont {Ala-Nissila}},\
  }\href@noop {} {\bibfield  {journal} {\bibinfo  {journal} {Phys. Rev. E}\
  }\textbf {\bibinfo {volume} {80}},\ \bibinfo {pages} {031602} (\bibinfo
  {year} {2009})}\BibitemShut {NoStop}%
\bibitem [{\citenamefont {Murnaghan}(1944)}]{MurnaghanFD:1944pnas}%
  \BibitemOpen
  \bibfield  {author} {\bibinfo {author} {\bibfnamefont {F.~D.}\ \bibnamefont
  {Murnaghan}},\ }\href@noop {} {\bibfield  {journal} {\bibinfo  {journal}
  {Proceedings of the National Academy of Sciences of America}\ }\textbf
  {\bibinfo {volume} {30}},\ \bibinfo {pages} {244} (\bibinfo {year}
  {1944})}\BibitemShut {NoStop}%
\bibitem [{\citenamefont {Birch}(1947)}]{BirchF:1947pr}%
  \BibitemOpen
  \bibfield  {author} {\bibinfo {author} {\bibfnamefont {F.}~\bibnamefont
  {Birch}},\ }\href@noop {} {\bibfield  {journal} {\bibinfo  {journal}
  {Physical Review}\ }\textbf {\bibinfo {volume} {71}},\ \bibinfo {pages} {809}
  (\bibinfo {year} {1947})}\BibitemShut {NoStop}%
\bibitem [{\citenamefont {Hohenberg}\ and\ \citenamefont
  {Kohn}(1964)}]{Hohenberg1964}%
  \BibitemOpen
  \bibfield  {author} {\bibinfo {author} {\bibfnamefont {P.}~\bibnamefont
  {Hohenberg}}\ and\ \bibinfo {author} {\bibfnamefont {W.}~\bibnamefont
  {Kohn}},\ }\href@noop {} {\bibfield  {journal} {\bibinfo  {journal} {Phys.
  Rev.}\ }\textbf {\bibinfo {volume} {136}},\ \bibinfo {pages} {B864} (\bibinfo
  {year} {1964})}\BibitemShut {NoStop}%
\bibitem [{\citenamefont {Kohn}\ and\ \citenamefont {Sham}(1965)}]{Kohn1965}%
  \BibitemOpen
  \bibfield  {author} {\bibinfo {author} {\bibfnamefont {W.}~\bibnamefont
  {Kohn}}\ and\ \bibinfo {author} {\bibfnamefont {L.~J.}\ \bibnamefont
  {Sham}},\ }\href@noop {} {\bibfield  {journal} {\bibinfo  {journal} {Phys.
  Rev.}\ }\textbf {\bibinfo {volume} {140}},\ \bibinfo {pages} {A1133}
  (\bibinfo {year} {1965})}\BibitemShut {NoStop}%
\bibitem [{\citenamefont {Kresse}\ and\ \citenamefont
  {Hafner}(1993)}]{Kresse1993}%
  \BibitemOpen
  \bibfield  {author} {\bibinfo {author} {\bibfnamefont {G.}~\bibnamefont
  {Kresse}}\ and\ \bibinfo {author} {\bibfnamefont {J.}~\bibnamefont
  {Hafner}},\ }\href {\doibase 10.1103/PhysRevB.47.558} {\bibfield  {journal}
  {\bibinfo  {journal} {Phys. Rev. B}\ }\textbf {\bibinfo {volume} {47}},\
  \bibinfo {pages} {558} (\bibinfo {year} {1993})}\BibitemShut {NoStop}%
\bibitem [{\citenamefont {Kresse}\ and\ \citenamefont
  {Furthm\"uller}(1996)}]{Kresse1996}%
  \BibitemOpen
  \bibfield  {author} {\bibinfo {author} {\bibfnamefont {G.}~\bibnamefont
  {Kresse}}\ and\ \bibinfo {author} {\bibfnamefont {J.}~\bibnamefont
  {Furthm\"uller}},\ }\href {\doibase 10.1103/PhysRevB.54.11169} {\bibfield
  {journal} {\bibinfo  {journal} {Phys. Rev. B}\ }\textbf {\bibinfo {volume}
  {54}},\ \bibinfo {pages} {11169} (\bibinfo {year} {1996})}\BibitemShut
  {NoStop}%
\bibitem [{\citenamefont {Perdew}\ \emph {et~al.}(1996)\citenamefont {Perdew},
  \citenamefont {Burke},\ and\ \citenamefont {Ernzerhof}}]{Perdew1996}%
  \BibitemOpen
  \bibfield  {author} {\bibinfo {author} {\bibfnamefont {J.~P.}\ \bibnamefont
  {Perdew}}, \bibinfo {author} {\bibfnamefont {K.}~\bibnamefont {Burke}}, \
  and\ \bibinfo {author} {\bibfnamefont {M.}~\bibnamefont {Ernzerhof}},\
  }\href@noop {} {\bibfield  {journal} {\bibinfo  {journal} {Phys. Rev. Lett.}\
  }\textbf {\bibinfo {volume} {77}},\ \bibinfo {pages} {3865} (\bibinfo {year}
  {1996})}\BibitemShut {NoStop}%
\bibitem [{\citenamefont {Bl\"ochl}(1994)}]{Bloechl1994}%
  \BibitemOpen
  \bibfield  {author} {\bibinfo {author} {\bibfnamefont {P.~E.}\ \bibnamefont
  {Bl\"ochl}},\ }\href {\doibase 10.1103/PhysRevB.50.17953} {\bibfield
  {journal} {\bibinfo  {journal} {Phys. Rev. B}\ }\textbf {\bibinfo {volume}
  {50}},\ \bibinfo {pages} {17953} (\bibinfo {year} {1994})}\BibitemShut
  {NoStop}%
\bibitem [{\citenamefont {Wu}\ \emph {et~al.}(2010)\citenamefont {Wu},
  \citenamefont {Adland},\ and\ \citenamefont {Karma}}]{Wu:2010vn}%
  \BibitemOpen
  \bibfield  {author} {\bibinfo {author} {\bibfnamefont {K.-A.}\ \bibnamefont
  {Wu}}, \bibinfo {author} {\bibfnamefont {A.}~\bibnamefont {Adland}}, \ and\
  \bibinfo {author} {\bibfnamefont {A.}~\bibnamefont {Karma}},\ }\href@noop {}
  {\bibfield  {journal} {\bibinfo  {journal} {Phys. Rev. E}\ }\textbf {\bibinfo
  {volume} {81}},\ \bibinfo {pages} {061601} (\bibinfo {year}
  {2010})}\BibitemShut {NoStop}%
\bibitem [{\citenamefont {Graham}(1998)}]{PhysRevLett.80.3888}%
  \BibitemOpen
  \bibfield  {author} {\bibinfo {author} {\bibfnamefont {R.}~\bibnamefont
  {Graham}},\ }\href {\doibase 10.1103/PhysRevLett.80.3888} {\bibfield
  {journal} {\bibinfo  {journal} {Phys. Rev. Lett.}\ }\textbf {\bibinfo
  {volume} {80}},\ \bibinfo {pages} {3888} (\bibinfo {year}
  {1998})}\BibitemShut {NoStop}%
\bibitem [{\citenamefont {Graham}(1996)}]{PhysRevLett.76.2185}%
  \BibitemOpen
  \bibfield  {author} {\bibinfo {author} {\bibfnamefont {R.}~\bibnamefont
  {Graham}},\ }\href {\doibase 10.1103/PhysRevLett.76.2185} {\bibfield
  {journal} {\bibinfo  {journal} {Phys. Rev. Lett.}\ }\textbf {\bibinfo
  {volume} {76}},\ \bibinfo {pages} {2185} (\bibinfo {year}
  {1996})}\BibitemShut {NoStop}%
\end{thebibliography}%

\end{document}